\documentclass[english,useAMS, usenatbib]{mn2e}
\usepackage[T1]{fontenc}
\usepackage[latin9]{inputenc}
\usepackage{units}
\usepackage{rotating}
\usepackage{url}
\usepackage{amsmath}
\usepackage{commath}
\usepackage{amssymb}
\usepackage{float}
\usepackage{graphicx}
\usepackage{esint}
\usepackage[authoryear]{natbib}
\usepackage{listings}
\usepackage{auto-pst-pdf}
\usepackage{color}

\makeatletter

\def\gtsima{$\; \buildrel > \over \sim \;$}
\def\ltsima{$\; \buildrel < \over \sim \;$}
\def\prosima{$\; \buildrel \propto \over \sim \;$}
\def\gsim{\lower.5ex\hbox{\gtsima}}
\def\lsim{\lower.5ex\hbox{\ltsima}}
\def\simgt{\lower.5ex\hbox{\gtsima}}
\def\simlt{\lower.5ex\hbox{\ltsima}}
\def\simpr{\lower.5ex\hbox{\prosima}}
\def\la{\lsim}
\def\ga{\gsim}

\providecommand{\tabularnewline}{\\}

\title[X-ray ionization of the IGM]{X-ray ionization of the intergalactic medium by quasars}
\author[Luca Graziani et al.]{Luca Graziani$^{1,2}$
\thanks{E-mail: luca.graziani@sns.it }, B. Ciardi$^{1}$, M. Glatzle$^{1,3}$\\
$^{1}$Max-Planck-Institut f\"ur Astrophysik, Karl-Schwarzschild-Stra{\ss}e 1, D-85748 Garching b. M\"unchen, Germany\\
$^{2}$Scuola Normale Superiore, Piazza dei Cavalieri 7, 56126 Pisa, Italy\\
$^{3}$Physik-Department, Technische Universit\"at M\"unchen, James-Franck-Str. 1,85748 Garching, Germany}
\usepackage{babel}
\usepackage{natbib}

\begin{document}

\date{Accepted 2010 <Month> XX. Received 2010 <Month> XX; in original form
2010 <Month> XX}

\maketitle
\pagerange{\pageref{firstpage}--\pageref{lastpage}} \pubyear{2010}\label{dat:firstpage}

\begin{abstract}

We investigate the impact of quasars on the ionization of the surrounding intergalactic medium (IGM) with the radiative transfer code \texttt{CRASH4}, now accounting for X-rays and secondary electrons. After comparing with analytic solutions, we post-process a cosmic volume ($ \approx 1.5\times 10^4$ Mpc$^3 h^{-3}$) containing a ULAS J1120+0641-like quasar (QSO) hosted by a $5 \times 10^{11} {\rm M}_\odot h^{-1}$ dark matter (DM) halo.  We find that: (i) the average HII region ($R\sim3.2$~pMpc in a lifetime $t_f = 10^7$~yrs) is mainly set by UV flux, in agreement with semi-analytic scaling relations; (ii) a largely neutral ($x_{\textrm{HII}} < 0.001$), warm ($T\sim 10^3$~K) tail extends up to few Mpc beyond the ionization front, as a result of the X-ray flux; (iii) LyC-opaque inhomogeneities induce a line of sight (LOS) scatter in $R$ as high as few physical Mpc, consistent with the DLA scenario proposed to explain the anomalous size of the ULAS J1120+0641 ionized region. On the other hand, with an ionization rate $\dot{N}_{\gamma,0} \sim 10^{57}$~s$^{-1}$, the assumed DLA clustering and gas opacity, only one LOS shows an HII region compatible with the observed one. We deduce that either the ionization rate of the QSO is at least one order of magnitude lower or the  ULAS J1120+0641  bright phase is shorter than $10^7$~yrs.

\end{abstract}


\section{INTRODUCTION\label{sec:INTRODUCTION}}

X-rays play a central role in astrophysics because they can penetrate the 
deepest layers of dense gaseous regions, inducing photo-chemistry 
and a complex chain of collisional ionizing processes (secondary ionization). Even in 
environments with a high column density of neutral hydrogen ($N_{\texttt{HI}} > 10^{20} \textrm{cm}^{-2}$), X-rays 
can create extended and partially ionized regions where a substantial fraction of the 
primary ionizing energy is converted into secondary, high velocity photo-electrons. 
These fast, charged particles are then able to collisionally ionize or excite both 
the H and He components of the gas. Since the first studies of Dalgarno and collaborators \citep{1963P&SS...11..463D}, 
X-rays have been recognized as a central agent regulating the gas chemistry in 
planetary  atmospheres, supernova ejecta, and the interstellar medium.

Quasi-stellar object  spectra, for example, show emission lines induced  
by X-rays entering the surrounding clouds of the host galaxy \citep{1984ApJ...280..269K,2010SSRv..157..177K} 
and easily escaping it into the IGM. 
These lines are an exceptional diagnostic tool of the physical state of the host galaxy 
ISM because they are very sensitive to temperature and gas electron density, and 
therefore they are impacted by secondary ionization.

Because of their large mean free path, X-rays can also leave a peculiar signature 
on the large scale of the intergalactic medium (IGM) by determining 
its temperature and ionization state \citep{2001ApJ...563....1V,2007MNRAS.375.1269Z,2008MNRAS.387..158R,2014Natur.506..197F, 2014ApJ...791..110X}. 
At high redshift even a small contribution by X-rays sources could produce an initial 
ionization of the IGM and increase its temperature \citep{2010A&A...523A...4B,2013MNRAS.431..621M,2017MNRAS.468.3718K, 2017MNRAS.471.3632M, 2017MNRAS.468.3785R, 2017ApJ...840...39M,2018MNRAS.476.1174E}  above the value of the Cosmic Microwave 
Background, affecting the detectability of the 21-cm signal from neutral hydrogen in these remote epochs \citep{1997ApJ...475..429M, 2003ApJ...596....1C,2007MNRAS.376.1680P,2014MNRAS.443..678P,2016MNRAS.460.4320E,2017ApJ...848...23K, 2017MNRAS.469.1166D, 2017MNRAS.472.4508S}.

The present generation of radio-array facilities such as the LOw Frequency ARray (LOFAR\footnote{www.lofar.org}), 
or the Mileura Wide-field Array (MWA\footnote{www.mwatelescope.org}), 
as well as the future Square Kilometre Array (SKA\footnote{www.skatelescope.org}), are expected 
to have the sensitivity required to detect the signal (e.g. \citealt{2007MNRAS.376.1680P}) before 
the epoch of hydrogen reionization. The interpretation of these observations will require a precise theoretical 
modeling of the ionization and heating processes.

Secondary ionization is important also in more exotic processes, as the decay or annihilation of 
Warm and Cold Dark Matter occurring during the Dark Ages. Such processes have 
been proposed in recent years as an additional source of IGM pre-heating 
(e.g. see \citealt{2008MNRAS.387L...8V, 2010MNRAS.404.1569V, 2012MNRAS.422..420E} and 
references therein).

Theoretical models of radiative transfer (RT) require an accurate implementation 
of both photo-ionization and secondary ionization to correctly interpret the data provided by
these new facilities, to increase the realism of both hydrogen and helium reionization simulations, 
and to make predictions of the temperature evolution of the intergalactic gas (e.g. \citealt{2001ApJ...553..499O,
2001ApJ...563....1V, 2003MNRAS.340..210G, 2010A&A...523A...4B, 2011MNRAS.417.2264V,2017MNRAS.468.3785R}).

Most of the photo-ionization codes in current active development 
\citep{1998PASP..110..761F,2008ApJS..175..534E,2008ApJS..178...20A} 
already account for the soft X-ray contribution and include the effects of the secondaries by implementing different models of energy release by cascade of charged particles. Fitting functions valid for H and He cosmological mixtures not polluted by metals have been provided 
by Shull and Van Steenberg \citep{1979ApJ...234..761S,1985ApJ...298..268S} and are currently 
implemented, among others, in \texttt{MAPPINGS III}, \texttt{CLOUDY} \citep{2013RMxAA..49..137F}, 
\texttt{MOCASSIN} and the semi-numerical \texttt{21CMFAST} code \citep{2011MNRAS.411..955M}.  
Cosmological radiative transfer codes, accounting for time evolution as well \citep{2002ApJ...575...33R,2010A&A...523A...4B, 2011MNRAS.414.3458W,2012MNRAS.421.2232F, 2014MNRAS.440.3778J, 2014ApJS..211...19B, 2016MNRAS.455.4406L} have also included the effects of secondaries, mainly preferring the Shull and Van Steenberg model with the prescriptions suggested by \cite{2002ApJ...575...33R}, but did not explore the dependence of their results on alternative or more updated models of secondaries (see Appendices A and B of the present work.).

In this work we first introduce a novel implementation of the cosmological radiative transfer code \texttt{CRASH} (hereafter \texttt{CRASH4}) which self-consistently treats UV and X-ray photons by accounting for the physics of secondary ionization. As various models of this process available in the literature treat the physical problem with different approximations (see Appendix A), \texttt{CRASH4} has been designed to embed any of them in a modular way, so that their impact on the code predictions can be easily assessed. Second, we benchmark the code predictions with a series of ideal tests investigating the impact of a bright quasar-like source on a surrounding homogeneous IGM; these test cases are essential to establish the average properties of both hydrogen and helium ionized regions.

Recent observations of the highest redshift quasars at $z\sim7.5$ \citep{2018Natur.553..473B} and $z\sim 7.1$ \citep{2011Natur.474..616M, 2011MNRAS.416L..70B} suggest that their surrounding IGM could be substantially neutral; the correct interpretation of their spectra requires a deep understanding of the structure and properties of their H$\,{\rm {\scriptstyle II}}$ regions and how they expand against collapsed structures present in the surrounding cosmic web. To this aim, we apply our code to a realistic case of a high redshift QSO surrounded by a neutral, well resolved IGM and we explore the evolution of the resulting ionized fronts as function of the source lifetime. The effects of both UV and X-rays photons on the IGM are then statistically analyzed by sampling the ionized volume with a large number of line-of-sights. Note that the results of the present work complement the findings of a companion paper (\citealt{2017MNRAS.468.3718K}, KK17), based on the same \texttt{CRASH4} code, but investigating the case of a higher redshift quasar  surrounded by star forming galaxies. KK17 suggests that galaxy surveys, performed on a region of sky imaged by 21 cm tomography, could help disentangling the relative role of QSOs and galaxies in driving cosmic reionization. This last problem is also addressed with \texttt{CRASH4} by performing large scale reionization simulations (see \citealt{2018MNRAS.476.1174E}).  


The present work is organized as follows. A general overview of \texttt{CRASH4} is provided in 
Section 2, where we also briefly introduce the soft X-ray band and the secondary 
ionization process. In Section 3, we provide an extensive series of tests to show 
the code reliability and flexibility in simulating different physical environments. 
A first application to a QSO environment is shown in Section 4
and the dependence of the results on the adopted model 
of secondaries is fully discussed in the related appendices.
The paper conclusions are summarized in Section 5.


\section{CRASH4 overview \label{sec:xrays}}

The aim of the new release of the radiative transfer code \texttt{CRASH}
\citep{2001MNRAS.324..381C,2003MNRAS.345..379M,
2009MNRAS.393..171M, 2013MNRAS.431..722G,2017MNRAS.467.2458H}, is 
the implementation of a new RT scheme capable to
simulate simultaneously the radiative transfer in many spectral bands: the 
Ly$\alpha$ spectral line, the Lyman-Werner band, the UV and soft X bands, and finally 
the energies up to the non-relativistic gamma rays limit.
In the present work we focus on the implementation of the soft X-ray band 
($E_{\gamma} \sim 200$~eV - 3~keV) and on the description of the secondary ionization models 
necessary to compute the gas ionization and temperature.

Since its first release, \texttt{CRASH} is able to simulate the hydrogen 
and helium photo-ionization induced by UV radiation ($E_{\gamma} \in [13.6-200]$~eV) 
emitted from point sources and/or a background. 
The emission scheme of the code has been progressively enhanced  
by introducing a better sampling of the spectral distribution, an accurate treatment 
of the gas temperature (see \citealt{2009MNRAS.393..171M}) and, finally, 
by treating the resonant propagation of $\textrm{Ly}\alpha$ photons self-consistently 
with the ionizing continuum \citep{2009MNRAS.393..872P}. 
The last global release of \texttt{CRASH} focused instead on implementing ionization of atomic metals 
self-consistently with the RT scheme \citep{2013MNRAS.431..722G} and the optional integration in galaxy formation hybrid pipelines \citep{2015MNRAS.449.3137G,2017MNRAS.469.1101G}. The introduction of metals, which are ionized both at energies below the H$\,{\rm {\scriptstyle I}}$ ionization potential 
(e.g. Si$\,{\rm {\scriptstyle I}}$) and at X-rays energies (e.g. O$\,{\rm {\scriptstyle VII}}$),
has highlighted the need for extending the RT to a wider spectral range. 

As several separate implementations of  \texttt{CRASH} already solve the RT in 
different energy bands (e.g. \citealt{2009MNRAS.393..872P}), \texttt{CRASH4} 
should be considered as the convergent release of the many code streams and has already 
allowed the extension of the RT scheme through dust polluted regions (Glatzle et al., submitted). 
The large set of new physical effects introduced by \texttt{CRASH4}, combined with 
a consistent and unified development of all its modules, provides more realism to 
reionization simulations accounting for different source types (KK17, \citealt{2018MNRAS.476.1174E}) and it is crucial to advance previous models of QSO regions (\citealt{2004MNRAS.350L..21M}, \citealt{2007MNRAS.376L..34M} (hereafter AM07), \citealt{2009MNRAS.395.1925M}), often limited to the investigation of hydrogen only I-fronts created by a UV flux.

The new multi-spectral-band RT scheme required a novel strategy 
capable of increasing the computational complexity while maintaining the 
flexibility of the algorithm. 
By increasing the code modularization and introducing many layers of  
parallelism, we implemented each spectral band in a separate threaded module. 
The ensemble of these new cooperative modules is organized into a new code 
library (\emph{Radiative Transfer Library for Cosmology}, hereafter \texttt{RT4C}). 
Each \texttt{RT4C} module also handles the new physics induced in the gas by 
photo-ionization in a specific spectral band (e.g. secondary ionization induced by X-rays) and provides 
the necessary extensions to the equations of ionization and thermal evolution. 

As more details on the algorithm can be found in previous {\tt CRASH} papers, 
here we just provide the basic information necessary to understand the 
details of the new implementation. 

A {\tt CRASH4} run consists of emitting photon packets from ionizing sources present in the  
computational domain and following their propagation through a Cartesian  grid of $N^{3}_c$ cells. 
At each cell crossing, \texttt{CRASH} evaluates the absorption probability for a 
single photon packet as $(1-e^{-\tau})$ and derives the corresponding number of 
photons absorbed in the cell as:
\begin{equation}
N_{abs}=N_\gamma(1-e^{-\tau}),
\label{eq:AbsPhot}
\end{equation}
where $N_\gamma$ is the photon content of the packet entering the cell and 
\begin{equation}
\tau= \sum_i n_i \sigma_{i} l,
\label{eq: TAU}
\end{equation}
is the gas optical depth in the cell due to the contribution of all chemical 
species $i$. $n_i$ is then the number density of species $i$, $\sigma_i$ is its 
photo-ionization cross section (PCS), and $l$ is the path casted in the cell by the crossing 
ray. For an easier reading, we have omitted the frequency dependency of 
$\sigma_i$, $\tau$ and $N_{abs}$ but it is important to note that the algorithm propagates 
photon packets (i.e. collection of photons ad different frequencies distributed 
according to the intrinsic source spectral shapes of the sources \citep{2009MNRAS.393..171M}). 
For this reason the method is intrinsically multi-frequency in each considered spectral band 
and all the above quantities have an explicit dependence on photon energies.
$N_{abs}$ is used to compute the contribution of photo-ionization and 
photo-heating to the evolution of the various species and the gas temperature $T$.
We remind here that the current ray tracing scheme relies on the infinite 
speed of light approximation (see Appendix D).

To properly extend the above scheme to the soft X-ray band, it is necessary to ensure that 
the PCS  adopted for each ionized species (e.g. H, He and metals ions) are coherently computed as function of the photon energies and maintain the required precision in all the spectral bands accounted for.
Note that this fundamental atomic data is generally provided as fitting functions combining  predictions of complementary theoretical methods constrained on experimental data. For example, \citet{1993ADNDT..55..233V,1995A&AS..109..125V} 
showed that their Hartree-Dirac-Slater calculations are very reliable  
for inner atomic shells while result less accurate at lower energies, especially for low ionized 
or neutral species. Calculations based on the higher level R-Matrix method \citep{1974CoPhC...8..149B, 1987JPhB...20.6379B, 2006AAMOP..54..237B} are available in the Opacity and IRON projects\footnote{http://cdsweb.u-strasbg.fr/topbase/home.html} \citep{1992RMxAA..23...19S, 2005MNRAS.360..458B, 2005MNRAS.362L...1S}) and provide more accurate 
predictions in the low energy regime, i.e. near the thresholds of atomic outer shells. 

It is then important for multi-frequency RT codes computing the ionization of astrophysical plasma, to rely on a coherent set of atomic PCS. 
{\tt CRASH4} adopts PCS for hydrogen and helium provided in \cite{1996ApJ...465..487V}
\footnote{The PCS are based on analytic fits for the ground states of atoms and 
all ions of the Opacity Project calculations (TOPbase version 0.7, \citet{1993A&A...275L...5C}) 
and are evaluated by interpolating and smoothing over atomic resonances and systematically compared with the available experimental data (see references in \citealt{1996ApJ...465..487V}).
For H-like and He-like species the fits show a correct non-relativistic asymptote and their maximum energy is set to $E_{max}=50$~keV. Note that all the fits are provided in different classes of accuracy: e.g. for hydrogen-like ions (H$\,{\rm {\scriptstyle II}}$, He$\,{\rm {\scriptstyle III}}$) they are accurate within 0.2\%, while the fit for 
He$\,{\rm {\scriptstyle I}}$ has an accuracy better than 2\% below $120$ eV, and better 
than 10\% in the higher-energy band. More details on the accuracy of this fits for other ions can be found in the original publication.}.

The inclusion of secondary ionization effects and heating by secondary electrons  
is described below and in Appendix A and B.
The electron released by a photo-ionization event at frequency $\nu$ and requiring a 
potential $E_{\textrm{th}}$, carries a kinetic energy 
$E_k = (h\nu - E_{\textrm{th}})$, which could be enough to induce 
ionization by collisions in the surrounding gas, as well as to 
excite and heat it. The efficiency of this process is generally modeled as a function 
of $E_k$ and of the pre-existing ionization state of the gas $x_e \equiv n_{\textrm{e}}/n_{\textrm{gas}}$, 
where $n_{\rm e}$ and $n_{\textrm{gas}}$ are the electron and gas number density, 
respectively\footnote{As a first order approximation, the dependence on the gas 
temperature $T$ is generally neglected, although in reality the electron would 
also thermalize with the gas (e.g. at $T\sim 10^4$ K the electrons with 
$E_k \le 1$~eV could cool down the gas); the precise heating should then 
be calculated also as a function of $T$.}. Recent numerical calculations provide 
tabulated values of the number of secondary ionizations per species $i$, $N_{s,i}(E_k, x_e)$, and of the 
fraction $f_Q(E_k,x_e)$ of the photo-electron energy $E_k$ contributing to the gas heating. 

For each species $i$, the ionization contribution from secondary electrons ($\delta x_i$) 
can be evaluated as  $\delta x_i \propto N_{abs} N_{s,i}(E_k, x_e)$, where 
$N_{abs}$ is defined in Eq. 1. 
Similarly, the contribution to the photo-heating of a cell can be evaluated as 
$\delta T= E_k f_Q(E_k,x_e)$.

In \texttt{CRASH4} we implement tables provided by three recent models of secondary ionization: \citet[hereafter SVS85]{1985ApJ...298..268S}, \citet[hereafter DG99]
{1999ApJS..125..237D} and \citet[hereafter VF08]{2008MNRAS.387L...8V}. While SVS85 
is the most widely used in the literature, DG99 has a more accurate treatment of some 
micro-physical processes and allows a future integration of the H2 component in the gas mixture. 
Finally, VF08 provides an updated implementations of atomic data present in the literature, 
and fitting formulas valid also in the hard X and gamma 
spectral bands \citep{2010MNRAS.404.1569V}. While all the above models are in global agreement for photon energies above 100~eV (see Appendix A), their different implementation makes them complementary. DG99, for example, explicitly treats the H2 component and photons with energies below 100~eV, and it is the ideal choice when the impact of both Lyman-Werner and ionizing bands is accounted for. VF08 provides instead an estimate of the Ly$\alpha$ re-emission triggered by high energy photons and should be selected when Ly$\alpha$ flux becomes a crucial aspect of the investigation. 

Before moving to a description of the results of this implementation, here we highlight that the conversion of 
the energy of secondary electrons into ionization is more efficient in a neutral or very poorly 
ionized gas, while as  $x_e$ increases, an increasingly larger 
fraction of the initial electron energy is converted into heating or excitation. We refer 
the reader to Appendix A for more details on the effects of secondary electrons, on the model-dependent 
function $N_{s,i}(E_k, x_e)$, and on assumptions of the different models implemented in {\tt CRASH4}. 

Finally, we note that, although the code allows for it, the radiation emitted from recombining gas
has not been explicitly followed in these tests; then the effect of the soft X-ray
component should be considered as a lower limit.


\section{Ideal ionized regions}

Here we present a series of tests designed to verify the correct implementation 
of the X-ray physics in {\tt CRASH4}, and the resulting implications. Appendices A, B and C 
provide analogous tests performed with different ionization models and 
in different pre-ionization configurations. Finally, the time evolution of 
these regions and the implications of adopting the infinite speed of light 
approximation are discussed in Appendix D.

We compute the properties of the ionized sphere created by 
a QSO-type, isotropic, point source embedded in a box of $110h^{-1}$ 
Mpc comoving (cMpc) at redshift $z=7.08$ (i.e. $\sim 19.0$~physical Mpc (pMpc or simply Mpc) by assuming $h=0.72$). The domain is mapped on a Cartesian grid of 
$N_c^3=512^3$ cells and the source is placed at the box corner (1,1,1). 
As basic set-up we consider a source with ionization rate $\dot{N}_{\gamma,0} =1.36 \times 10^{56}$ 
photons s$^{-1}$ and a power-law spectral shape $S(\nu) \propto \nu^{-1.5}$ \footnote{Note that a single power-law index is often inappropriate for sources extending from the UV to the X-rays band \citep{2002ApJ...565..773T, 2005A&A...432...15P} but is generally adopted in models investigating high redshift ionized regions \citep{2011MNRAS.416L..70B}. We defer this model improvement to future investigations when a direct comparison with X-rays  observations will be performed.}. 
The source spectrum extends up to $E_{\textrm{max}}= 3$~keV and  it is sampled by $N_f = 41$ 
frequency bins, which are reduced if a cut to lower energies is applied (see next section for more details).
The source emission is  sampled by  $N_p = 2 \times 10^8 $ photon packets throughout the 
duration of the simulation, $t_f=10^8$~yrs. This choice ensures a Monte Carlo convergent 
solution as verified by repeating the run with a number 
of packets increased by one order of magnitude.
The gas is assumed to be homogeneous (with a number density 
$n_{\textrm{gas}}=9.59 \times 10^{-5}$ cm$^{-3}$), neutral, and at an initial 
temperature $T_0=100$~K. All the values indicated above have been selected in light 
of a direct comparison with the realistic case investigated in Section 4 and consistent with  
parameters adopted in AM07 and \citealt{2011MNRAS.416L..70B}.

\subsection{Hydrogen only}

In this section we describe the H$\,{\rm {\scriptstyle II}}$ region created in a  
gas composed by atomic hydrogen only. 
For this set-up we choose DG99 because it is the only 
model describing a configuration with pure H, while  SVS85 and VF08 
are designed to account for a cosmological gas mixture of H and He. 

The spherically averaged profile of $x_{\textrm{HII}}$ at the intermediate 
time $t_i=6\times10^7$~yrs is shown in the top panel of Figure 1, as obtained by cutting the assigned spectral shape 
up to different values of $E_{\gamma}$: $0.1$~keV (i.e. $E_{\gamma} \in [13.6-100]$~eV, dashed-double-dotted red line), 0.2~keV (dashed-dotted red line), $E_{\gamma} = 0.5$~keV (triple-dotted red line) and finally  a full case including photons with energies up to 3~keV (solid blue line). All the profiles account for the contribution of direct photo-ionization and secondary processes. Finally a case with maximum energy $E_{\gamma} = 3.0$~keV but excluding  the contribution of secondary ionization is shown as dashed black line\footnote{We remind the reader that for energies below 100~eV secondaries follow a less efficient, non linear regime (see Appendix A).}.
In each case the location of the I-front\footnote{Here we define the I-front as the 
distance at which $x_{\textrm{HII}} $ drops below 0.9. This definition, though, is somewhat arbitrary and a
different one might be more appropriate depending on the shape of the front and the problem at hand.}
remains the same and corresponds to $d \sim 2.0$ physical Mpc, indicating that 
the UV radiation below $E_{\gamma} = 0.1$~keV plays the most important role in establishing the gas full ionization\footnote{Note that to capture the time evolution of the entire profile the selected configuration does not have the correct spatial resolution for a comparison with semi-analytic predictions. A comparison is provided in Section 4 on a smaller box size (25h$^{-1}$~cMpc) and in Appendix D.}.

Photons with $0.1$~keV $< E_{\gamma} < 0.2$~keV induce instead a substantial difference 
in the external part of the H$\,{\rm {\scriptstyle II}}$ region (2.0~Mpc $ < d < 4.0$~Mpc), while the soft X-ray component is mainly supporting the tail at distances $d > 4.0$~Mpc. Their highly penetrating flux creates then a large and diffuse low-ionization shell 
with $0.0001 < x_{\textrm{HII}} < 0.003$. As a reference, the comoving mean free path $\lambda_X$ of an X-ray photon with $E_{\gamma} = 1$~keV in a H-only neutral IGM ($\bar{x}_{\textrm{HI}} \sim 1$) at $z=7.08$, is $\lambda_X = 4.9 \times \bar{x}_{\textrm{HI}}^{-1/3}((1+z)/15)^{-2} (E_{\gamma}/300~{\rm eV})^{3} = 625$~ cMpc, i.e. 77 pMpc (see \citealt{2006PhR...433..181F,2012RPPh...75h6901P}). At $d > 2.0$~Mpc, the 
efficiency of secondary electrons in ionizing the gas is also evident, 
increasing the ionization fractions of about one order of magnitude compared to a case without
secondaries {and maintaining the extreme tail of the profile up to $d\sim 8$~Mpc}. It is important to note here that by disabling secondary ionization the code removes this effect at all photon energies, as 
imposed by the model in DG99. This is the reason why the dashed black line (i.e. without secondaries) 
corresponds to values lower than the dashed-dotted red line.
It is also evident, by comparing the 0.2~keV and 0.1~keV cases, that part of the photons with $E_{\gamma} > 0.1$~keV contributes in creating the external tail discussed above, as the efficiency of the  secondary ionization process has no activation thresholds but increases with the photon energy (see Appendix A for more details).

\begin{figure}
\centering
\includegraphics[angle=-90,width=0.50\textwidth]{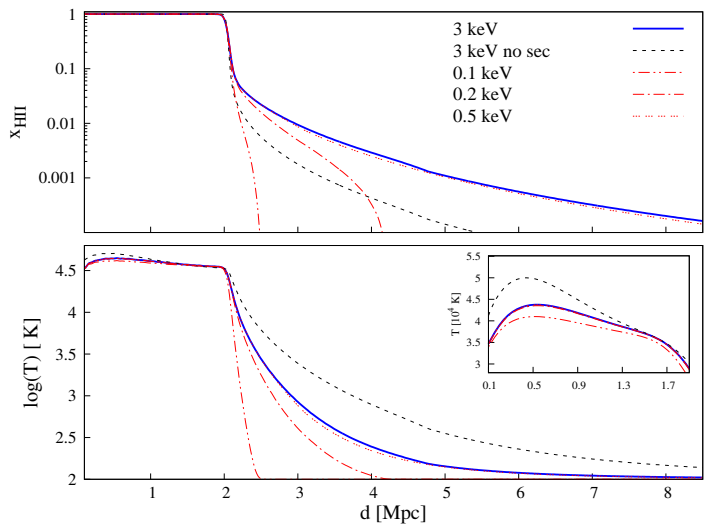}
\vspace{1truecm}
\caption{Radial profiles of $x_{\textrm{HII}}$ (top panel) and log$(T)$ (bottom) of the Str\"omgren sphere created in a gas made of 
         neutral atomic hydrogen at the simulation time  
         $t_i=6\times10^7$ yrs, and adopting the DG99 model. The distance $d$ from the source is shown 
         in physical Mpc. In all the panels the profiles are created by 
         a spectrum extending up to: $E_{\textrm{max}}= 3$~keV, with/without the effect of the secondary 
         electrons (blue solid/black dashed line).
         The dashed-double-dotted red line indicates the profile created by cutting 
         the energy at $E_{\textrm{max}}= 0.1$~keV, the dashed-dotted red at $E_{\textrm{max}}= 0.2$~keV, 
         while the triple-dotted red line 
         refers to a cut at $E_{\textrm{max}}= 0.5$~keV. 
         The small box in the bottom panel shows a zoom of the gas temperature (in linear scale
         of $10^4$ K) in the fully ionized region 0.1 Mpc $< d <1.9$ Mpc.
         }
\end{figure}

The bottom panel of Figure 1 shows the corresponding temperature profiles. Similarly 
to $x_{\rm HII}$, also the temperature inside the I-front 
($T \sim 4.0 \times 10^4$ K) is mainly determined by the effect of photons with $E_{\gamma} < 0.1$~keV,
while higher energy photons in the spectrum induce a remarkable 
gas heating in the external shell ($2.0$~Mpc$ < d < 2.9$~Mpc), where $T \ge 10^3$K. 
A zoom-in view of the radial distribution of $T$ in $0.1$~Mpc $ < d < 1.9$~Mpc is given in the panel inset, where we show the 
temperature values in linear scale to better understand the contribution 
of secondaries in subtracting energy to the gas photo-heating.
Also note that even at larger distance (corresponding to $d > 3.5$~Mpc) 
the temperature is maintained at values above $T_0=100$~K.

When secondary ionization is switched off, 
all the kinetic energy of the fast electrons goes into gas heating  
rather than being distributed between heating and ionization; the resulting 
gas temperature is then higher than in the case with secondaries\footnote{It should be noted that while there are many ways to "disable" the effects of secondary ionization, either by conserving the total energy or not, each set-up will result in un-physical ionization fractions and/or temperature. The case with "no-secondary ionization" is then discussed here only to provide an estimate of its effect on global ionization vs. direct photo-ionization.}.
By comparing the red lines, the contribution to the gas heating by photons with $E_{\gamma} < 0.2$~keV becomes evident (also see panel inset).

\subsection{Hydrogen and helium}

We now discuss the ionized spheres obtained in a gas composed by hydrogen 
(number fraction 0.92) and helium (number fraction 0.08). We adopted the DG99 model
because it is the only one explicitly accounting for the effects of secondary 
ionization on $x_{\rm HeIII}$.
A discussion of the differences induced by using the other models (SVS85 and VF08) 
can be found in Appendix B.

\begin{figure}
\centering
\includegraphics[angle=-90,width=0.50\textwidth]{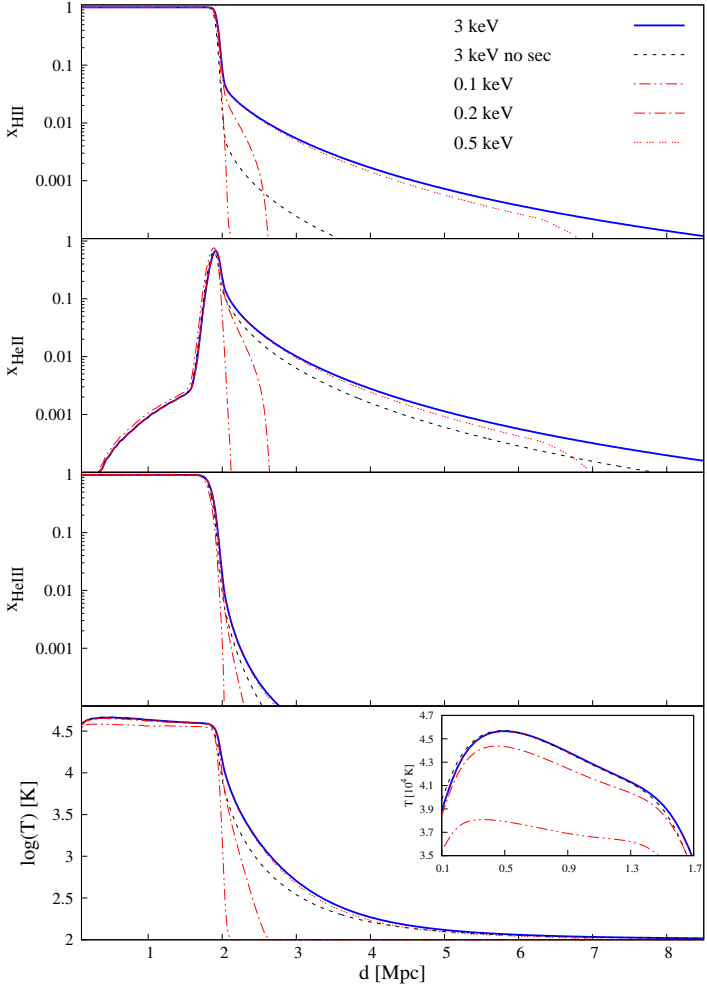}
\vspace{1truecm}
\caption{Radial profiles of the Str\"omgren sphere created in 
         a cosmological mixture of H and He at the simulation 
         time $t_i=6\times10^7$ yrs, and adopting the DG99 model.
         The distance $d$ from the source is shown in physical Mpc. 
         In all the panels the profiles are created by 
         a spectrum extending up to: $E_{\textrm{max}}= 3$~keV, with/without the effect of the secondary 
         electrons (blue solid/black dashed line).
         The dashed-double-dotted red line indicates the profile created by cutting 
         the energy at $E_{\textrm{max}}= 0.1$~keV, the dashed-dotted red at $E_{\textrm{max}}= 0.2$~keV, 
         while the triple-dotted red line 
         refers to a cut at $E_{\textrm{max}}= 0.5$~keV. 
         From the top to the bottom the panels refer to the profiles of 
         $x_{\textrm{HII}}$, $x_{\textrm{HeII}}$, $x_{\textrm{HeIII}}$ and log$(T)$.
         The small box in the bottom panel shows a zoom of the gas temperature (in linear scale
         of $10^4$ K) in the fully ionized region $0.1 < d <1.7$ Mpc.
         See text for more details. 
         }         
\end{figure}

In Figure 2 we show the results of this test. The location of the I-front of $x_{\textrm{HII}}$ is similar to the pure H case, indicating that helium does not substantially alter the ionization of the inner region.
On the other hand, the partially ionized shell at $d > 2.0$~Mpc created by photons with energy $E_{\gamma} \in [0.1-0.2]$~keV is 
reduced in size because He$\,{\rm {\scriptstyle I}}$ and 
He$\,{\rm {\scriptstyle II}}$ provide an important contribution
to the optical depth even at the lowest energy cuts.

The qualitative agreement found with Figure 1 indicates that the presence of secondary electrons primarily affects the hydrogen component, which is more abundant and can easily support low hydrogen ionization $x_{\textrm{HII}} < 0.001$ up to 8~Mpc. 

The second and third panels from the top show the profiles of $x_{\textrm{HeII}}$ 
and $x_{\textrm{HeIII}}$, respectively. The location of the I-fronts of 
$x_{\textrm{HeII}}$ and the size of the partially ionized shells are similar 
to those of $x_{\rm HII}$. This result depends on the spectral shape index 
chosen for this test.    
By comparing the solid and dashed lines in the first three panels, we can confirm 
the dominant role of the secondaries in maintaining the external shell at low 
ionization, although their effect is less significant than for hydrogen.
Neutral He is then more sensitive to direct photo-ionization by X-rays photons than to 
secondary ionization.
The X-ray component has an impact also on the $x_{\textrm{HeIII}}$, 
although smaller than on the other profiles. A little role (similar 
to what is found for $x_{\rm HeII}$) is also played by the secondaries in maintaining 
the partial ionization level of the outer region (compare dashed and solid lines).  

The bottom panel finally shows the temperature profiles. 
Here again, the major differences induced by the presence of X-rays are found 
in the shell outside the I-front: while the inner region is maintained at a 
temperature $T \sim 4.2 \times 10^4$~K , the 
outer shell has temperatures in the range  $10^3$~K$ < T < 10^4$~K up to a distance of 
$d \sim 2.5$ Mpc, i.e. the end of the $x_{\textrm{HeIII}}$ profile. The initial temperature 
$T_0 = 100$ K is finally restored after a long and smoothly decreasing tail corresponding to the decrease of the $x_{\textrm{HII}}$ ionization profile. 
A zoom of the temperature behavior in the fully ionized region is provided also for this configuration. 
Here the contribution of photons with $0.1$~keV $< E_{\gamma} < 0.5$~keV is more evident as they mainly interact with the helium component (compare solid and dashed-dotted/dashed-double-dotted lines). Photons with energies $E_{\gamma} > 0.5$~keV do not affect the heating of the inner region as their profile overlaps with the solid blue line.
Finally, note that when helium is included the energy is distributed in a more complex way across the three ions. For example, we note that the temperature of the low ionization tail obtained by disabling the secondary ionization (black dashed line) is lower than in the case with X-rays (solid blue) and does not have a straightforward interpretation. Here  two effects combine: (i) the He column density is lower than the H one, and thus the collisional ionization probability on atoms of helium results one order of magnitude lower (see blue lines in Figures A1, A2, Appendix A), 
and (ii) at X-ray frequencies the photo-ionization cross section of neutral helium is about one order of magnitude higher  
than the one of hydrogen.   

We have repeated the test presented above in a range of gas number density 
($n_{gas} \in [10^{-2}, 10^{-5}]$~cm$^{-3}$), source ionization rates ($\dot{N}_{\gamma,0} \in [10^{54} -10^{57}]$ ~photons s$^{-1}$) 
and non neutral initial conditions $x_0$ ($x_{\textrm{0,HII}}=x_{\textrm{0,HeII}}=x_0 \in [10^{-4}- 2\times10^{-1}]$ 
and $x_{\textrm{0,HeIII}}=0$) to verify that the qualitative results illustrated above are not numerical 
artifacts nor features restricted to a specific set-up. As expected, we found a similar behavior
in all configurations. Quantitatively though, there are differences with e.g. the size of the partially ionized shell strongly depending on 
$x_0$ as detailed in Appendix C\footnote{The dependence on $T_0$ is negligible in all the cases.}. 
On the other hand, the location of the I-front remains the same as long as $x_0<0.1$, while it progressively moves away from the source with increasing $x_0$. In realistic cases, involving many sources and an in-homogeneous gas distribution, the problem becomes strongly model-dependent.

To summarize the results of this section, the presence of X-rays creates an extended region at low ionization and temperatures up to $10^{3}$ K, which is not present at photon energies $E_{\gamma} < 0.1$~keV. This feature is confirmed by all the models of secondary ionization (see Appendix B), even if the internal structure of the ionized shell can significantly change from one model to another. 

\section{IGM ionization by quasars}

In this section we consider a QSO-like source at $z \sim 7.08$ embedded in a massive halo, in analogy with \citet{2011MNRAS.416L..70B}. The gas distribution is simulated using the code {\tt GADGET3} \citep{2005MNRAS.364.1105S} in a box of comoving size 25~Mpc~$h^{-1}$, chosen so that the most massive halo at $z \sim 7.08$ 
(with a total mass of $\sim 5 \times 10^{11}$~M$_\odot h^{-1}$, $R_{vir}\sim 292.2$~ckpc $\sim 36.2$~kpc physical) is located in a corner of the box\footnote{We use the following cosmology: $\Omega_m$=0.26, $\Omega_\Lambda$=0.74, $\Omega_b$=0.0463, $h$=0.72, $\sigma_8$=0.85 and $n$=0.95.}. The simulation employs $2 \times 512^3$ gas and dark matter particles, resulting in a particle mass resolution of $1.50 \times 10^6 {\rm M}_\odot  h^{-1}$ and 
$6.91 \times 10^6 {\rm M}_\odot  h^{-1}$, respectively, and a gravitational softening length of 1.628~kpc$~h^{-1}$.
The outputs of this hydrodynamical simulation are mapped onto a grid of $N_c^3=512^3$ cells and post-processed with {\tt CRASH4}. The resulting spatial resolution is $d \sim 48.8 h^{-1}$ ckpc, i.e.  $\sim 8.4$ physical kpc. 

We place our source within the most massive halo located in the grid corner (1,1,1), and consider the emission in a quadrant angle.
The gas mixture is composed by hydrogen and helium in cosmological fractions and we only consider the case of an initial neutral gas configuration\footnote{The initial temperature in each cell is taken from the hydro simulation. We have verified that the neutral set-up is consistent with the average thermal properties of the box because of the absence of strong shocks and significant initial collisional ionization.}; this set-up allows us to understand the contribution of the gas inhomogeneities in enhancing or suppressing the features found in the test cases.
In the reference run the source emissivity is $\dot{N}_{\gamma,0} = 1.36 \times 10^{56}$ photons s$^{-1}$ as in the ideal tests to allow a direct comparison. In addition, we also investigate the implications of a higher value $\dot{N}_{\gamma,1} = 1.36 \times 10^{57}$ photons s$^{-1}$, consistent with the ionization rate estimated for the high redshift quasar ULAS J1120+0641 \citep{2011Natur.474..616M}\footnote{\citet{2011MNRAS.416L..70B} already noted that observations of lower redshift QSOs suggest that this value should be associated with a higher DM halo mass, many uncertainties still persist about the effects of QSO clustering or the exact  association of quasars with cosmic web over-densities. To allow a more straightforward comparison with other theoretical works we also consider the set of parameters indicated in the text to be more consistent with these studies.}, while the other parameters are the same as in the previous section. The effects of the two ionization rates in a uniform medium are discussed in Appendix B (see also \citealt{2015MNRAS.454..681K}).

For this study we adopt a full spectrum including X-rays with energies up 3~keV and the VF08 model of secondary ionization, providing the most updated 
treatment of the atomic coefficients and the most recent modeling of collisional processes induced 
by secondaries (see Appendix A and B for more information).
 The simulation starts assuming the gas as neutral and takes as initial temperature $\bar{T}_0$ the average value of $\sim 100$~K provided by the gas hydrodynamics. The simulation duration is set up to $t_f= 10^8$~yrs, with results stored at intervals of $\Delta t= 10^7$~yrs, consistent with an average quasar lifetime (see \citealt{2016ApJ...824..133K} for a recent estimate at lower redshift). This set-up significantly improves a series of previous studies (e.g. AM07) both in the spatial resolution, in the accuracy of the RT physics and in the derived properties of the IGM (i.e. its ionization and temperature). It should be noted that old treatments of RT were computationally limited to a shorter UV range (AM07 uses $E_{\gamma} \in [13.6 - 42]$~eV) appropriate for computations with hydrogen only, while {\tt CRASH4} is designed to overcome these limitations and to provide a more realistic description of the multi-frequency RT through cosmic mixtures.  
As final remark we point out that this set-up does not consider any previous reionization history; we defer the discussion of this important point to the companion work KK17. 

\subsection{Average properties of the ionized region}

To understand the impact of the quasar on the simulated volume, we first summarize the average properties of the HII region  at an intermediate time $t_i=6\times10^7$~yrs and at the end of the simulation $t_f=10^8$~yrs\footnote{More details on the correct choice of the right simulation time can be found in Appendix D.}. 
The volume average hydrogen ionization fractions are $\bar{x}_{\textrm{HII}} \sim 0.162$ and $\bar{x}_{\textrm{HII}} \sim 0.274$, with a minimum value in the simulated  volume of about $x_{\textrm{HII}} \sim 3 \times 10^{-4}$ in both cases. These values clearly indicate that (i) the fully ionized region involves only $\sim 15 $\% (30 \%) of the simulated volume and its impact on the ionization of the neutral IGM is limited in space\footnote{The real impact of QSOs in the IGM would require a detailed analysis of their role in the context of full reionization simulations including the effects of the galaxy UV background. This point is introduced in KK17. Also see Appendix C below, where we investigate the sensitivity of our results to gas ionization.}, (ii) the X-ray flux pervades the entire cosmological box impacting gas far from the fully ionized fronts. Similar considerations can be applied to the helium component, which results fully ionized in 12 \% (20 \%) of the volume and is never completely neutral, clearly indicating the presence of high energy photons at the end of the simulated volume. Consistently with these findings, the volume-averaged gas temperature is almost in photo-ionization regime, $\bar{T} \sim 5.5\times10^{3}$ $(\sim 8.5\times10^{3})$ K, with a large spread from  $T_{\textrm{max}} \sim 4.0 \times 10^{4}$ $(\sim 8.4 \times 10^{4})$ K down to  $T_{\textrm{min}} \sim 115$ K (a negligible heating for the scope of the present work). The spherically-averaged profiles of the ionized region are similar to the profiles found in the previous section, predicting an I-front position around $d \sim 2.61$ $(\sim 3.2)$ physical Mpc, which is consistent with semi-analytic estimates of a sphere radius $R_s$ (derived by \citealt{2000ApJ...542L..75C, 2000ApJ...542L..69M} and further discussed in \citealt{2003AJ....126....1W}):

\begin{equation}
R_s =7.0\mathrm{\,pMpc}\left(\frac{\dot{N}_{\gamma,0}}{10^{58}\mathrm{s^{-1}}}\frac{t_{s}}{10^{7}\mathrm{yr}}\right)^{\nicefrac{1}{3}}\left(\frac{1+z}{7.37}\right)^{-1},
\end{equation}
yielding $R_s = 2.77$ $(3.3)$ physical Mpc, when computed for values appropriate to our simulation\footnote{Note that the sphere size is also consistent with the AM07 estimates by adopting the values of our configuration, but a more extended spectrum and a full configuration with H and He. Additional comparisons with other semi-analytic models and RT codes can be found in KK17.}. These estimates, though, can not account for the effects of X-rays outside the I-front and can not describe the large oscillations revealed along different LOSs. These are discussed in the next section.      

\subsection{Direction-dependent ionized region size}

To analyze the impact of the cosmic web on the structure of ionization fronts, we draw $10^6$ random LOS from the central source; we also checked that their casted path captures both the hydrogen and helium I-fronts and the long partially ionized tail. We start discussing the longest/shortest LOS (LOS A/LOS B) I-front found at the reference time $t_i=6\times10^7$~yrs and at the end of the simulation, while the statistical properties of the sample are presented at the end of the section\footnote{The equivalent ideal ionized sphere is discussed in Appendix B and can be used for a comparison.}.

\begin{figure}
\centering
\includegraphics[angle=-90,width=0.50\textwidth]{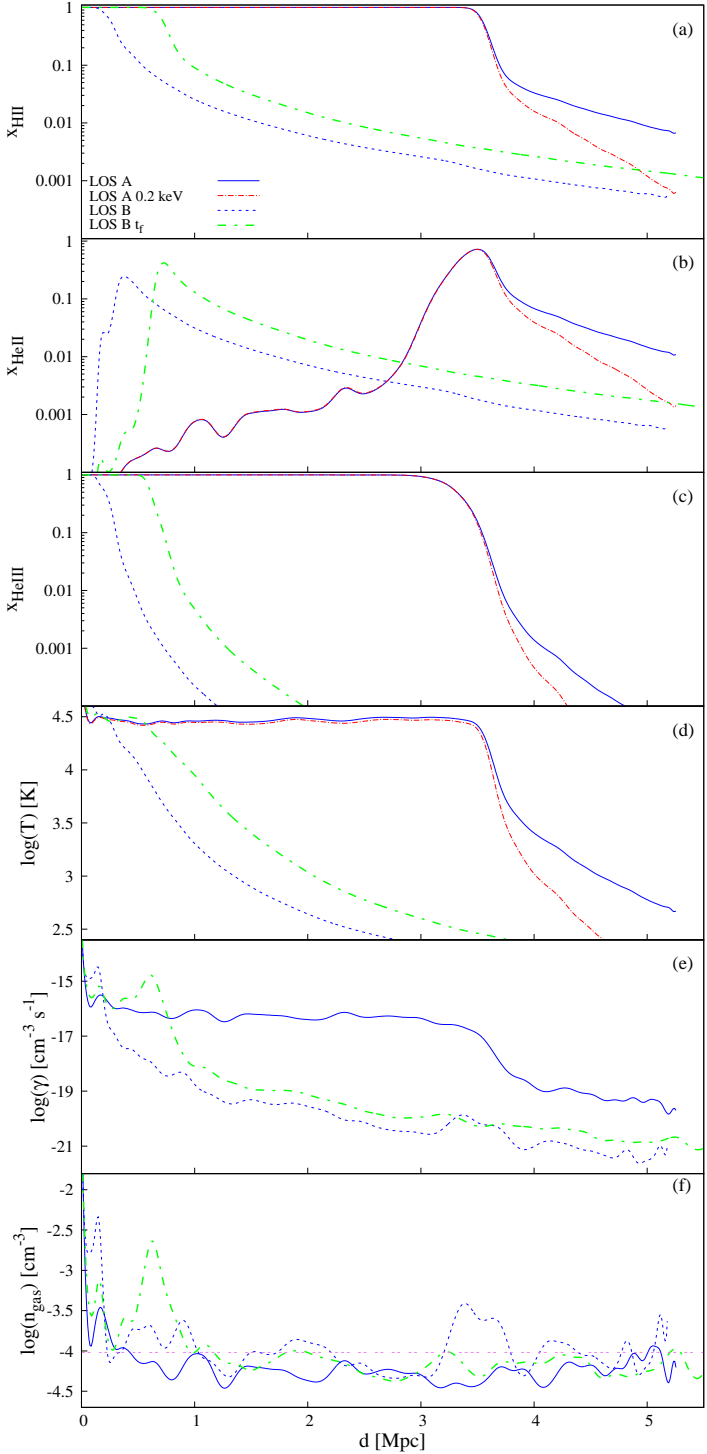}
\vspace{1truecm}
\caption{The most (LOS A, blue solid lines) and the least (LOS B, blue dashed lines) 
         extended ionization profile in the realistic ionized region created around 
         the QSO considered in Section 4. Red lines with same linestyle refer to the 
         same LOS profiles derived from a simulation performed with a spectrum cut at 0.2~keV.
         The lines refer to the simulation time $t_i= 6\times10^7$ yrs and 
         the distance $d$ from the source is shown in physical Mpc. The green dashed-dotted line refers instead to the LOS B (UV+X) found in the same simulation at the final time $t_f=10^8$~yrs.
         From the top to the bottom the panels refer to the profiles of $x_{\textrm{HII}}$, 
         $x_{\textrm{HeII}}$, $x_{\textrm{HeIII}}$, log$(T)$, log$(\gamma)$ and log$(n_{gas})$.
         The horizontal line in the bottom panel indicates the volume average value of the gas number density 
         in the adopted hydro-dynamical snapshot.
}
\end{figure}

Figure 3 shows LOS A/LOS B in both $E_{\gamma} = 3.0$~keV and $E_{\gamma} = 0.2$~keV  spectrum configurations, together with the corresponding profiles of temperature, gas recombination rate ($\gamma$) and number density\footnote{Note that to capture the entire spatial extension, here the line has been smoothed with a bezier interpolation suppressing small scale numerical fluctuations induced mainly by the gridding of SPH particles when deriving density and temperature fields. An accurate study of  real fluctuations in $x$ and $T$ induced by the RT at kpc scales would require a higher spatial and mass resolution as well as a better Monte Carlo sampling of the radiation field at the I-front. We defer this investigation to future studies exploiting the new AMR features of \texttt{CRASH}, as introduced in \citet{2017MNRAS.467.2458H}. Finally, note that the over-density peaks shown in Figure 3, as well as their implications on the radial fluctuations of the HII region, are not affected by the bezier smoothing.}. 
It is immediately evident that the position of the hydrogen I-fronts of the two LOSs differs by $-92$\% and $+26$\% from the semi-analytic estimate $R_s\sim 2.77$. These differences are mainly due to the high number density present in the inner region of the halo (corresponding to over-densities $\Delta = n_{\textrm{gas}} / \bar{n}_{\textrm{gas}} >100$) and the various gas structures found along each LOS, as seen in panel (f). 
Note that  the global shape of the profiles is broadly similar to that of spherically averaged profiles (see Appendix B and compare with Figure B1), while significant spatial $x_{\textrm{HeII}}$ oscillations are induced by the underlying density distribution.

At $t=6\times10^7$~yrs the I-front of LOS B is trapped within $d \sim 0.2$ Mpc, as photons with energies lower than 0.2~keV cannot escape the inner over-density (for this reason its line is not visible in the plot), while X-rays can travel up to end of the box and create a smoothly declining profile at low ionization ($0.0001 < x_{\textrm{HII}} < 0.01$), which extends up to $d\sim 5$~Mpc.
A similar interpretation applies to LOS A, intercepting a smoother gas distribution (mostly under-dense) and consequently finding its I-front pushed away by $\sim3.5$ physical Mpc. 

These differences are mainly due to photons $E_{\gamma} < 0.2$~keV (UV regime), while the X-rays photons create an extended layer of ionized hydrogen at $0.1 < x_{\textrm{HII}} < 0.001$ with extension $d\sim 4.5$ Mpc, physical. For both LOSs, the external region of the box is characterized by values $x_{\textrm{HII}} \leq 10^{-3}$. As in the ideal tests (refer for e.g. to  Figure 2), low ionization is produced by secondary electrons, whose effects are modulated by variations of the recombination rate associated with gas density peaks (see  panel (e)).

Similarly to hydrogen, the $x_{\textrm{HeII}}$ profiles become more extended due to X-rays, while the fully ionized helium is hardly affected, as it depends mainly on the UV flux.

The temperature profiles are shown in panel (d) of Figure 3. Note, for LOS A, the presence of a long tail at $T \sim 10^3$ K extending up to $d \sim 4.5$ Mpc, and temperatures as high as $T \sim 500 $ K found up to distance $d \sim 6$ Mpc; beyond this distance the gas has not been reached by photons and its temperature is at $\bar{T}_0$. This is a clear indication that the X-ray heating in the region surrounding the QSO is not negligible. The gas temperature of LOS B follows the same trend and decreases below $T \sim 10^4$~K at $d>0.5$~Mpc. Note, on the other hand that it remains above $T=10^3$~K up to $d \sim 1.5$~Mpc indicating substantial heating by X-ray photons. The initial temperature is finally restored at distances $d>2.5$~Mpc.

We performed the previous analysis at the simulation end ($t_f=10^8$~yrs) to investigate the front displacement, finding a similar relative scatter ($\sim 4$~Mpc). LOS B at $t_f$ shows that UV photons are stopped by over-densities ($\Delta > 30$) present within 1~Mpc (see Figure 3, dashed-dotted green line). Depending on the assumed QSO lifetime, systems clustering around the QSO can then significantly shape its I-front and change its impact on the surrounding IGM (see also KK17). More quantitatively, we find that 0.5 \% of our I-fronts stop within 2~Mpc while only $\sim9$ \% within 3~Mpc. Despite their effectiveness in trapping the UV flux, the probability of observing a radius smaller than $R_s$ is then very low, at least within the adopted clustering. 

For a better comparison with the case of ULAS J1120+0641, in the following we investigate how the least extended LOS changes by adopting $\dot{N}_{\gamma,1} = 1.36 \times 10^{57}$~s$^{-1}$ and a lifetime $t_i=10^7$~yrs. In this case LOS B shows an I-front at $d \sim 1.7$ Mpc and it is particularly remarkable as it is created by a sudden increase of $n_{\textrm{gas}}$ corresponding to a series of collapsed structures with gas over-density $\Delta > 100$. In the left panel of Figure 4 we show the cosmic web in a slice cut intercepting this region (dashed white square). By visualizing the gas distribution we verified that there is an intricate structure in which many filaments cross. Many systems with column densities $19  \leq \log (N_{\textrm{gas}}/{\rm cm}^{-2}) < 21$ are found in this region. The I-front of LOS B can be easily identified as the feature traced by the iso-contour lines of $x_{\textrm{HII}}$ in decreasing value, from red (0.99996) to blue (0.015). A zoomed-in view of the region of interest is given in the four panels on the right side of the same figure. Here the gas ionization  profiles of both hydrogen and helium are shown, as well as the gas temperature. The simultaneous presence of DLA ($N_{\textrm{gas}} \sim 4\times 10^{20}$~cm$^{-2}$ as well as many Lyman limit systems (i.e. more than 35 systems with $19.9  \leq \log (N_{\textrm{gas}}/{\rm cm}^{-2}) \leq 20.3$ in the cubic region relative to the zoomed-view) causes the front to drop sharply both in $x_{\textrm{HII}}$ and $T$. Consistently, the fully ionized helium front is found at shorter distances, being more sensitive to the gas over-densities via recombinations. This picture is remarkably consistent with the QSO properties observed by \citet{2011Natur.474..616M} and the DLA scenario proposed in \citet{2011MNRAS.416L..70B} for the interpretation of the HII region size\footnote{Note that the measured radius of the ionized near zone of ULAS J1120+0641 is three times smaller than a typical QSO in $6.4 < z < 6$ and, in absence of a contaminating DLA, the IGM neutral fraction estimated at $z\sim7.1$ would be a factor of 15 higher than the one at $z\sim6.2$.}. The spectrum of ULAS J1120+0641, exhibits in fact a Ly$\alpha$ damping wing which could be produced by both a substantially neutral IGM in front of the QSO, and the presence of a DLA along the observed LOS (although the number of these systems clustering around a QSO is generally very small). \citet{2011MNRAS.416L..70B} investigated the implications of the above scenarios by simulating 100 Ly$\alpha$ absorption spectra produced by a high-resolution hydrodynamical simulation and a 1D radiative transfer code. The authors conclude that the derived transmission profiles could be consistent with an almost neutral IGM in the vicinity the QSO with an assumed lifetime $t_i \in [10^6 - 10^7]$~yr. On the other hand, if a DLA lies within 5 proper Mpc of the quasar, the possibility of an highly ionized IGM cannot be ruled out (but note that this only occurs in $\sim5$\% of their LOSs). Finally, a QSO bright phase longer than $10^7$~yr would imply, for their model, a too large near-zone.

Despite the above similarities, our simulation shows that the number of lines of sight found within 2~Mpc at $t_i=10^7$~yrs is only 0.02 \% of the entire sample. At successive times (from $t=3\times10^7$~yrs) the over-dense systems get fully ionized and the ionized region occupies more than 40 \% of the volume. At late times (i.e. from $t=6\times10^7$~yrs) more than 90 \% becomes fully ionized. In Figure 5 we show the same region at the final time of the reference set-up (i.e. $\dot{N}_{\gamma,0}$, $t_f=10^8$~yrs). Note that while the inner region is still fully ionized (although with a slightly lower ionization fraction and temperature) the shaping of the I-front is remarkably similar to the previous case and the DLA discussed above is still capable to stop the ionizing flux. 

Quasar lifetime and ionization rate are both crucial to determine the impact of the QSO on the surrounding IGM, against over-densities. Within the clustering of DLA systems predicted by the adopted hydro-dynamical simulation, our radiative transfer run clearly shows that in a pre-assumed neutral IGM, the probability to find a LOS showing an ionized region smaller than 2~Mpc is extremely low. As long as the quasar lifetime is sufficiently long (i.e. larger than $6\times10^7$~yrs), this remains valid even with the lowest ionization rate. In the brightest case (corresponding to the ionization rate in \citealt{2011Natur.474..616M}) the duration of the QSO optically bright phase is instead strictly limited below $\sim 10^7$~yrs. 

\begin{figure*}
\centering
\includegraphics[scale=0.7,angle=0]{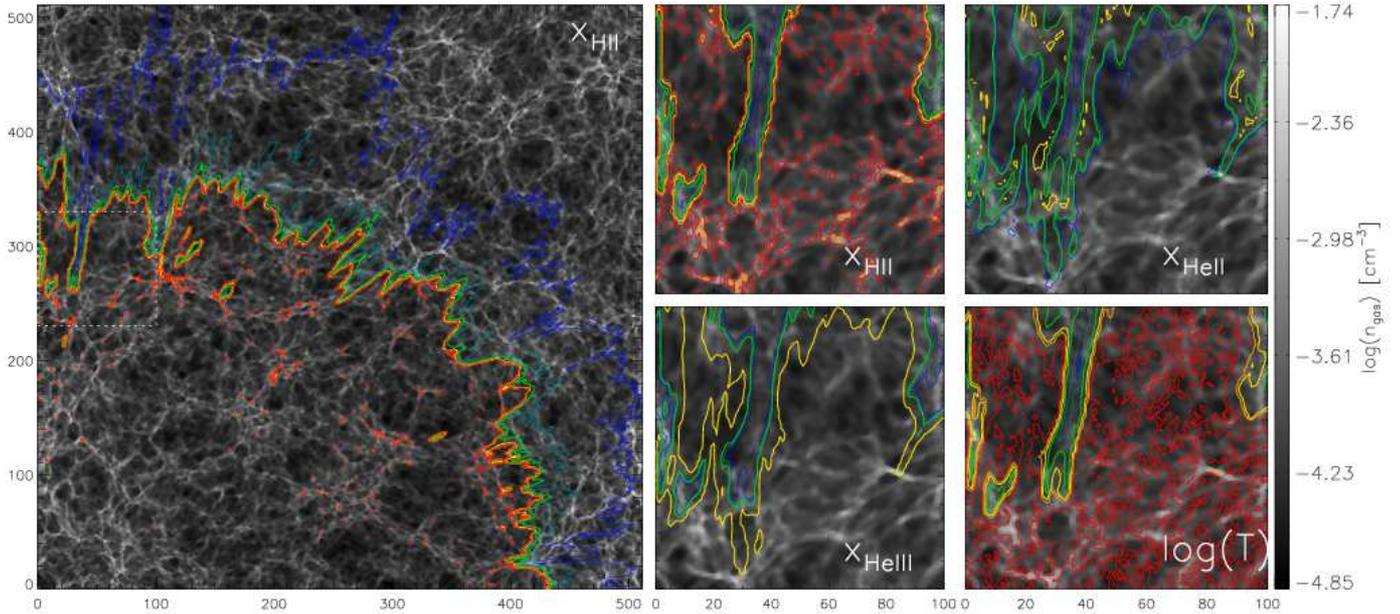}
\vspace{-0.3truecm}
\caption{Slice cut of the box intercepting the region containing LOS B for a simulation with  $\dot{N}_{\gamma,0} = 1.36 \times 10^{57}$~s$^{-1}$ at $t=10^7$~yrs. $n_{\textrm{gas}}$ is shown as gray color palette, while the iso-contour lines represent fronts at progressively lower ionization fraction, $x_{\textrm{HII}} = {0.99996, 0.99995, 0.9, 0.1, 0.05, 0.015}$, from red to blue. The box dimensions are shown in cell units (1 cell $\sim 0.05h^{-1}$~cMpc $\sim 8.4$~kpc
physical), as a reference the virial radius of the halo is contained in the first four cells. The region crossed by LOS B in Figure 3, is indicated by the dashed square and zoomed-in in the four panels on the right. In these panels iso-contour lines of $x_{\textrm{HII}}$ (top left), $x_{\textrm{HeII}}$ (top right), $x_{\textrm{HeIII}}$ (bottom left) are shown with the same color scheme. The bottom right panel adopts the same color scheme to map iso-contour regions corresponding to $\log(T/ {\rm K}) = [4.55, 4.4, 4.0, 3.5, 3.0, 2.69]$.
}
\end{figure*}

\begin{figure*}
\centering
\includegraphics[scale=0.7, angle=0]{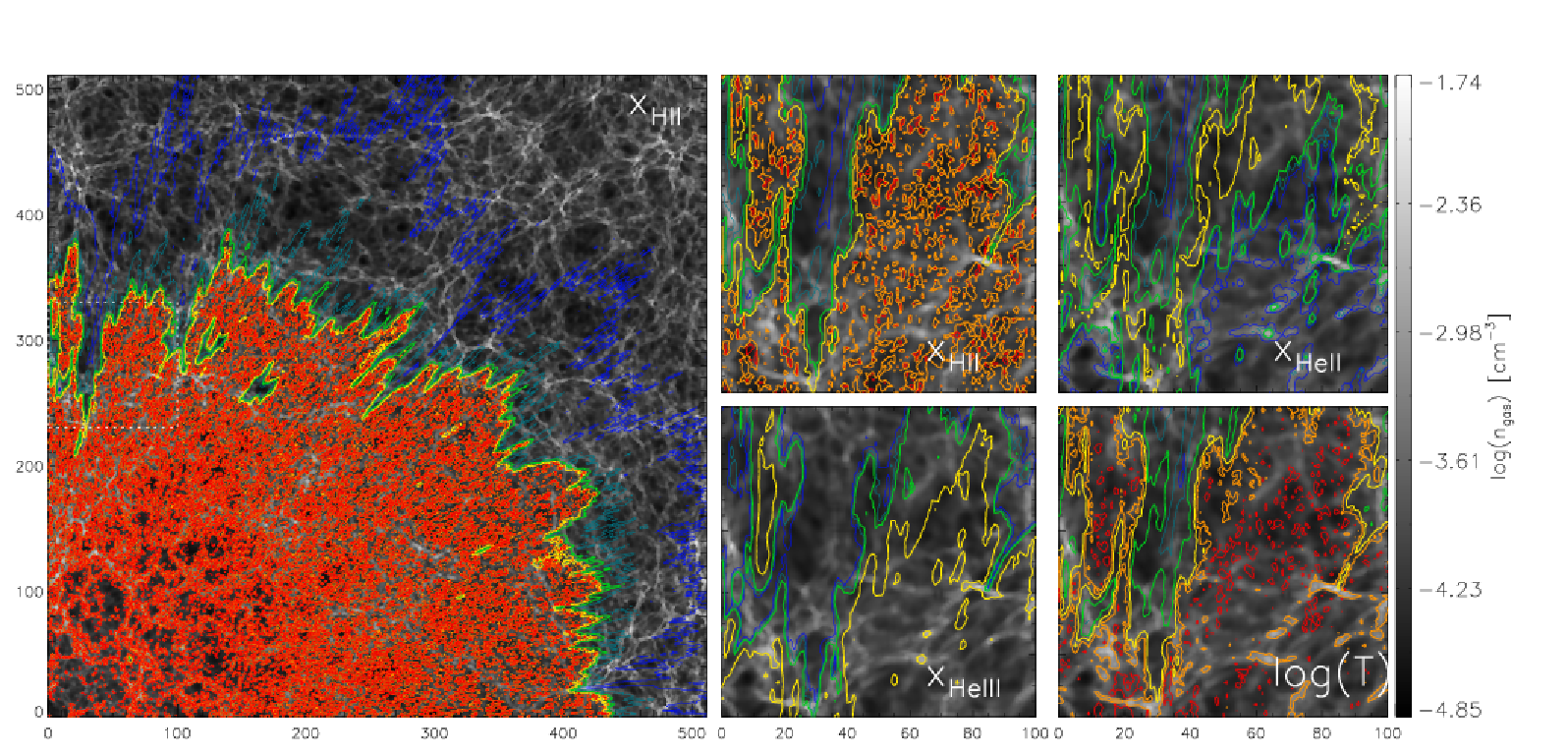}
\vspace{-0.5truecm}
\caption{Slice cut of the box intercepting the region containing LOS B for a simulation with  $\dot{N}_{\gamma,0} = 1.36 \times 10^{56}$~s$^{-1}$ at $t_f=10^8$~yrs. $n_{\textrm{gas}}$ is shown as gray color palette, while the iso-contour lines represent fronts at progressively lower ionization fraction, $x_{\textrm{HII}} = {0.99996, 0.99995, 0.9, 0.1, 0.05, 0.015}$, from red to blue. The box dimensions are shown in cell units (1 cell $\sim0.05h^{-1}$~cMpc $\sim 8.4$~kpc
physical), as a reference the virial radius of the halo is contained in the first four cells. The region crossed by LOS B in Figure 3, is indicated by the dashed square and zoomed-in in the four panels on the right. In these panels iso-contour lines of $x_{\textrm{HII}}$ (top left), $x_{\textrm{HeII}}$ (top right), $x_{\textrm{HeIII}}$ (bottom left) are shown with the same color scheme. The bottom right panel adopts the same color scheme to map iso-contour regions corresponding to $\log(T/ {\rm K}) = [4.55, 4.4, 4.0, 3.5, 3.0, 2.69]$.
}
\end{figure*}

\subsection{Statistic of ionization fronts}

We now turn to a statistical analysis of the distribution of $d_{max}$, defined as the maximum distance after which the ionization fraction is found always below an assigned threshold value $x_{th}$
\footnote{This condition is always guaranteed \textit{against} the oscillations discussed in Section 4.1 because (i) there is only one source with a decreasing flux along $d$ and (ii) it is embedded in a neutral medium which always guarantees zero $x$ and low $T$ at domain boundaries.}. 
Guided by the ideal cases discussed in Section 3 and Appendix B and C,  we investigate the values $x_{th}= 0.001, 0.01, 0.1, 0.9$, descriptive of X-rays or UV (i.e. photons with $E_{\gamma} \leq 0.2$~keV) dominated regions found in the reference case at $t=6\times10^7$~yrs\footnote{Note that this choice guarantees a stable I-front fully contained in our simulated volume. See Appendix D for more details.}.

\begin{figure}
\centering
\includegraphics[angle=-90,width=0.5\textwidth]{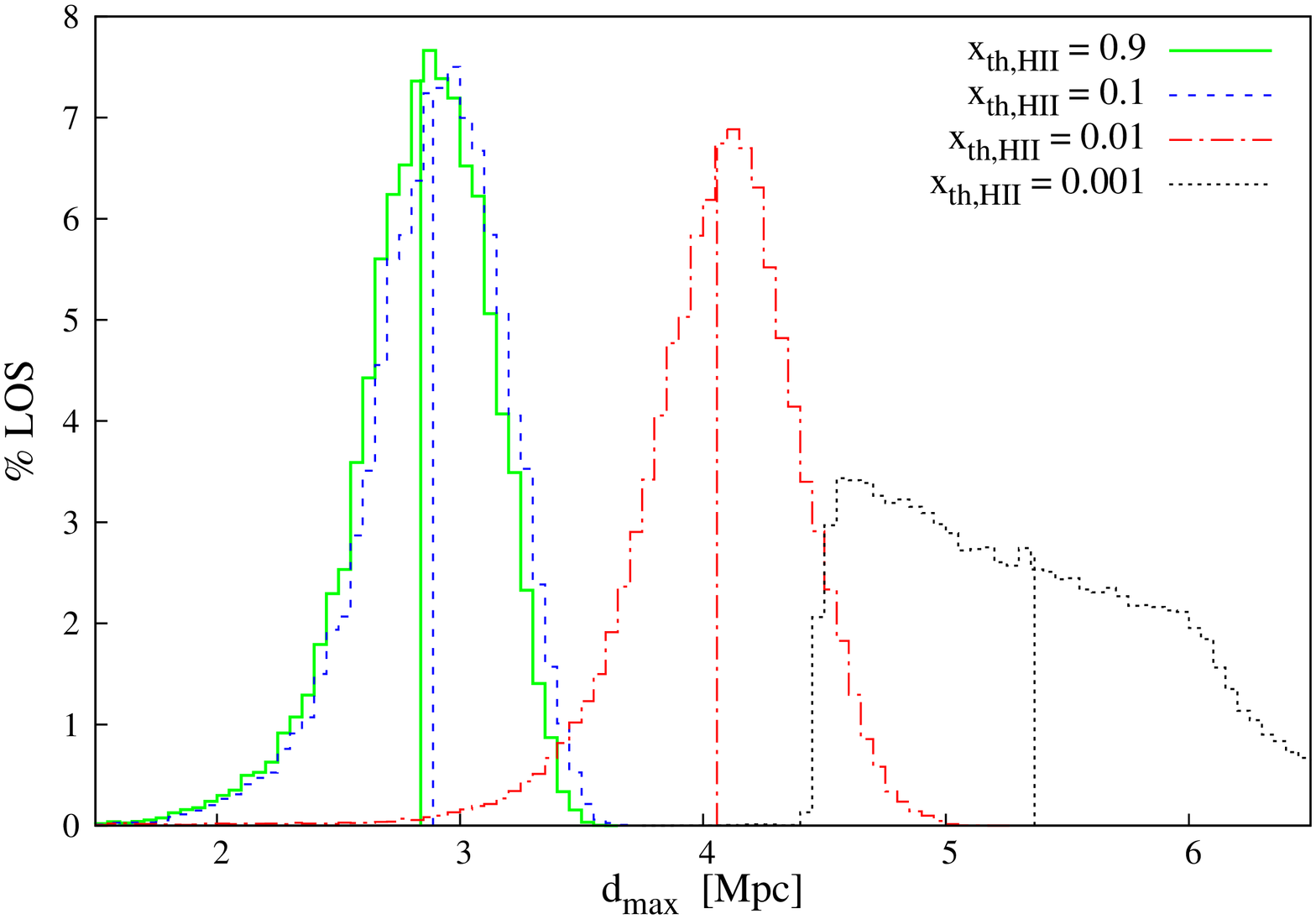}
\caption{Percentage of LOS showing the maximum distance $d_{max}$ [physical Mpc] at which $x_{\textrm{th,HII}}$ = 0.9 (solid green line), 0.1 (dashed blue line), 0.01 (dashed-dotted red line), and 0.001 (dotted black line). Vertical lines show the average value of the distance, $\bar{d}_{max}$, with the same color conventions. The statistic refers to the simulation time $t_i=6\times10^7$ yrs.}
\end{figure} 

In Figure 6 we show the percentage of LOS for which a given $d_{max}$ value is found for a given $x_{\textrm{th,HII}}$\footnote{Note that each distribution is normalized with respect to its own entries and not to the total number of LOSs.}. 
First note that the solid green and the dashed blue histograms are very similar and have average values of $\bar{d}_{max}= 2.84$ and 2.89~Mpc\footnote{Note that the value found for the ideal case of reference in Appendix B is $d \sim 2.1$~Mpc.}. 
We associate these distributions with the I-front set up by the UV flux: the similar shape of the histograms and their 
close average values are the statistical indication of an abrupt decline of the I-front. The width of the 
histogram indicates instead the spread of $d_{max}$ : when $0.1 < x_{\textrm{HII}} <0.9$, only $\sim 60$\% of these LOS are found in 2.7~Mpc $< d_{max}<3.2$ Mpc. 
The tails of the distributions indicate that less than 0.1\% of LOS are found in $d_{max}<2.7$~Mpc and in $d_{max}>3.2$~Mpc.

The X-rays dominated regions are instead associated with $x_{\textrm{th,HII}}=0.01$ and $x_{\textrm{th,HII}}=0.001$. The average values of their distributions are far apart from those of the UV I-front, of about 1.2 and 2.5~Mpc, i.e $\bar{d}_{max} \sim 4.1$ Mpc and $\bar{d}_{max} \sim 5.4$ Mpc, respectively. Also, note the extended width ($\approx$ 1-2 Mpc) of these low ionization regions.

To confirm the physical interpretation provided above, we now investigate the statistic of $d_{max}$ for different values of $x_{\textrm{th,HeIII}}$. In this case we expect a significant overlap in the histograms of $d_{max}$ at various ionization fractions because He$\,{\rm {\scriptstyle III}}$ does not show any extended ionization layer in the ideal tests. This statement is statistically verified in Figure 7, which has the same line and color organization of the previous one. First note the overlap of the histograms and how close their average values are (vertical lines): only $d \sim 0.7 $~Mpc separate the solid green and dotted black lines; this is the statistical indication of a compact ionization front, rapidly decreasing beyond the UV dominated front ($x_{\textrm{th,HeIII}} = 0.9$ at $\bar{d}_{max} \sim 2.8$~Mpc). As commented in Section 3.1.2, the low ionization rate of X-rays beyond the UV I-front is not sufficient to sustain fully ionized helium.

The same analysis (not shown here for brevity) has been performed for He$\,{\rm {\scriptstyle II}}$ and does confirm the previous comparative results.

\begin{figure}
\centering
\includegraphics[angle=-90,width=0.5\textwidth]{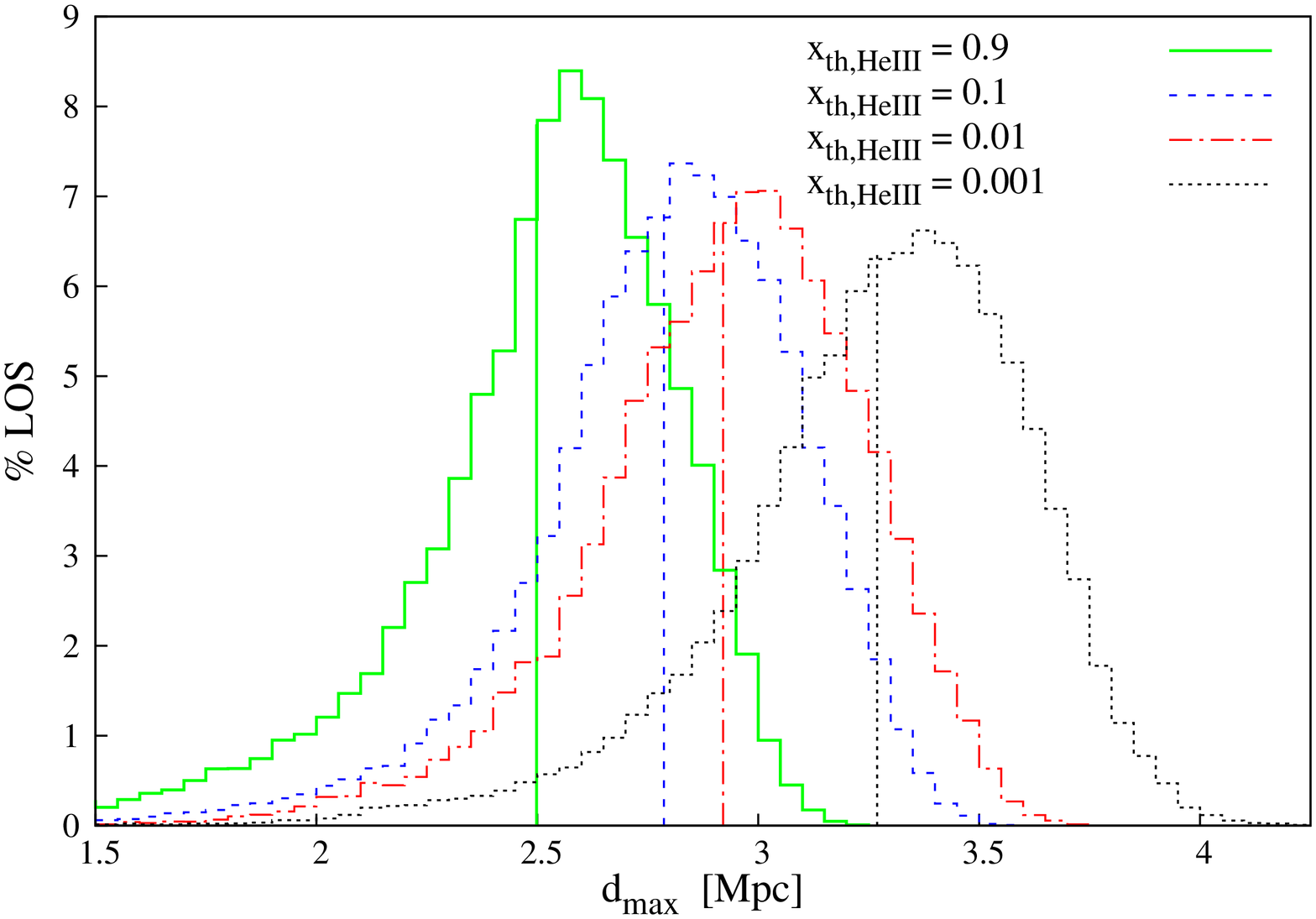}
\caption{Percentage of LOS showing the maximum distance $d_{max}$ [physical Mpc] at which $x_{\textrm{th,HeIII}} = 0.9$ (solid green line), 0.1 (dashed blue line), 0.01 (dashed-dotted red line), and 0.001 (dotted black line). Vertical lines show the average value of the distance $\bar{d}_{max}$, with same color conventions. The statistic refers to the simulation time $t_i=6\times10^7$ yrs.}
\end{figure} 

We conclude from  the statistic of ionization fronts that also with realistic gas distributions X-rays create an extended, highly-fluctuating low ionization layer up to few  physical Mpc, far from the emitting source.

\section{CONCLUSIONS}

This paper describes the impact of X-rays emitted by a high-redshift quasar 
on the surrounding IGM by using detailed RT simulations. The simulations are
performed with the new release of the cosmological radiative transfer code 
\texttt{CRASH4} which treats the RT problem in multi-frequency bands and 
self-consistently computes the gas temperature. After a 
brief introduction of the new code architecture,  we focus on the X-ray band 
implementation. We first describe the various secondary ionization models implemented in 
\texttt{CRASH4}, pointing out the importance of a correct modeling of secondary 
ionization process. The correctness of the numerical implementation and 
the complications introduced by the X-ray band are first investigated in a series of 
idealized tests by discussing the effects of an X-ray flux on a 
pure hydrogen case and on a cosmological mixture of hydrogen and helium.

As a first astrophysical application, we focus on a realistic case of a quasar-like source targeting 
the high redshift quasar ULAS J1120+0641 \citep{2011Natur.474..616M}. In this initial investigation we assume 
that the surrounding IGM is neutral when the QSO powers on.  Under this 
hypothesis we study the detailed structure of its ionization fronts. 
From an accurate statistical analysis of the front along a large number of lines of sight
we draw the following conclusions.

\begin{itemize}  

\item The main ionization pattern observed in the region surrounding the QSO is determined 
      by its UV flux, while the specific signature of X-rays consists in the creation of an extended
      low-ionization layer, highly sensitive to the gas over-densities surrounding the quasar halo.

\item When statistically analyzed through many lines of sight, the quasar ionization front shows an 
      average size consistent with semi-analytic estimates found in the literature, but also exhibits a 
      non negligible scatter higher than 2 physical Mpc (i.e  from $-92$\% to $+26$\% from the 
      average value $R_s$), mainly induced by the presence of an X-ray flux in the QSO spectrum. Our 
      statistical analysis also shows that the presence of a DLA system along a specific line of sight 
      can significantly bias the estimate of the HII region radius, and consequently the deduced 
      progress of reionization through proximity effect. Further investigations will be performed with 
      the same gas distribution to assess the importance of radiative transfer effects in presence of 
      a pre-ionized IGM, extending the findings of KK17.

\item Within the observed ionization rates and the adopted DLA clustering from the hydro-dynamical 
      simulation, we find only one LOS showing an HII region comparable with the observed one. By 
      comparing the region created by different ionization rates and quasar lifetimes we deduce that 
      either the ionization rate of the QSO is at least one order of magnitude lower or its bright 
      phase is limited to $t_f \ll 10^7$~yrs to support a DLA scenario with a minimum probability.
      
\end{itemize}
 
The cosmological consequences on recombination lines, temperature history of the IGM, on the progress of helium reionization and on the detectability of the 21-cm signal will be investigated with future high-resolution reionization simulations. 

\section*{ACKNOWLEDGMENTS}

The authors are grateful to the anonymous referee for his very constructive comments. We thank J. Bolton for providing the hydrodynamical simulations used in this paper, and to A. Ferrara, M. Shull, C. Evoli and M. Valdes for enlightening discussions. LG acknowledges the support of the DFG Priority Program 1573 and partial support from the European 
Research Council under the European Union's Seventh Framework Programme (FP/2007-2013) / ERC Grant Agreement n. 306476.


\bibliographystyle{mn2e}
\bibliography{XRaysIon}

\begin{appendix}

\section{Secondary ionization models}
The physics of secondary ionization was first dealt with by Spencer and Fano by providing the 
degradation spectrum of electrons and by calculating the energy losses 
induced by fast electrons \citep{1954PhRv...93.1172S}. More recent calculations account 
for the effects of gas ionization and metallicity: simulations in a partially ionized medium 
have been carried out by many authors and generally include sophisticated treatments of 
the micro-physical processes involving gas mixtures of H, He and eventually H2 
\citep{1968ApJ...152..971S,1971ApJ...166..525H, 1973A&A....25....1B,1979ApJ...234..761S,
1985ApJ...298..268S,1991ApJ...375..190X,1999ApJS..125..237D, 2008MNRAS.387L...8V}.
The last four implementations are briefly reviewed in 
the following; the interested reader can find more details in the cited literature. 

Here we describe the secondary ionization models 
implemented in \texttt{CRASH4} by briefly reviewing the main features and 
constraints of these models. Hereafter the quantities $f_{Q}$ and $f_{x_{i}}$ will refer to the 
fraction of the kinetic energy of the electrons, $E_k$,  that goes into gas 
heating and collisional ionization of species $i$, respectively.

Different methods have been used to evaluate the efficiency of secondary ionization and heating,
which depends on both $E_k$ and the ionization state of the gas, $x_e\equiv n_{\textrm{e}}/n_{\textrm{gas}}$, where $n_{\rm e}$ and $n_{\textrm{gas}}$ are the electron and gas number density.

Early studies employed the continuous slowing 
down approximation (see \citealt{1985JaJAP..24.1070P} and references therein), while 
more recent investigations took into account the discrete nature of energy loss processes,
either by solving the Spencer-Fano equation \citep{1954PhRv...93.1172S} or by performing 
Monte Carlo numerical integrations \citep{1971P&SS...19.1653D,1973A&A....25....1B,
1979ApJ...234..761S}. The reliability of these results is determined largely by the accuracy of 
the cross sections adopted for the micro-physics of the excitation, ionization and dissociation 
processes.

While the efficiency of the secondary ionization process depends non-linearly on $E_k$ and $x_e$ for energies $<$100~eV, at $E_k>$100~eV the energy distribution of a primary electron remains approximately constant and the efficiency of the process has a linear dependence on $E_k$.  

In gas mixtures it is generally assumed that the secondary electrons lose energy mainly in collisions 
with hydrogen because it normally dominates in mass. On the other hand, the presence of helium cannot 
be neglected because its cross sections are larger at the resonant 
photo-ionization potentials. When the gas is composed by H, He, H2 and metals, the process 
becomes increasingly more difficult to model and the definition of gas ionization fraction 
must be adapted to the various species releasing electrons into the gas by collisions. 

In the following we will discuss in more detail the three models implemented in {\tt CRASH4}.

\subsection{The DG99 model} 

DG99 \citep{1999ApJS..125..237D} has an accurate treatment of many micro-physical processes including the presence of a molecular component in the gas mixture. 
In fact, this model simulates the slowing down of fast electrons in a 
partially ionized gas composed of H, He and H2. The energy deposition is 
calculated by a Monte Carlo scheme following the history of an electron 
with initial energy $E_k$ \citep{1971ApJ...166..525H} through the total 
number of excitations, ionizations and dissociations.
The model includes many processes of excitation of atomic hydrogen and 
helium and also accounts for double ionization of helium. 
Although the energy loss due to electron impact excitation and ionization of  
He$\,{\rm {\scriptstyle II}}$ is small, the excitation of the resonance 
line at $30.4$ nm is accounted for because of its effects on the secondaries. 

Fittings functions are given for five energy intervals in the energy range $E_k=30$~eV -1~keV with an accuracy, limited by the bin size, of about 2\%.
The fits are valid only for $10^{-6} \leq x \leq 0.1$, where $x$ is the ionization fraction. 
In the range of energies and ionization fraction of common validity, the DG99 numbers are in good agreement with those from SVS85, the discrepancies being mainly due to updated atomic data and a more accurate treatment of atomic processes. 


Because in DG99 values of $f_{x_i}$\footnote{To be precise, DG99 provide the mean energy per ion pair, $W$, which can be converted into number of collisional ionization as $N_{s_i}=E_k/W_i$.} and $f_Q$ are provided for different compositions of the gas, in {\tt CRASH4} we have implemented tables corresponding to the five energy bins for a case of pure H, as well as for a gas composed of H and He in cosmological abundance. The energy interval is extended from 1~keV to 3~keV by linear extrapolation.
Following the approach of \citet{2005A&A...436..397M}, for a fixed value of $E_k$, the tables can be extended to $x>0.1$ using the analytic functions provided in  \citet{1991ApJ...375..190X}.  For $x<10^{-6}$ we assume instead that the gas is fully neutral.
The original binning in $x$ and $E_k$ are maintained as provided in the paper, while 
intermediate values are obtained by linear interpolation. 

\subsection{The SVS85 model}
 
Even if outdated in the atomic coefficients, SVS85 has been adopted by all the codes 
participating in the Cosmological Radiative Transfer Project \citep{2006MNRAS.371.1057I,
2009MNRAS.400.1283I} and it is included in \texttt{CRASH4} because it results useful for code 
comparison and testing.

The model is based on a Monte Carlo simulation \citep{1979ApJ...234..761S} which follows the 
degradation of the primary electron energy through various interactions with H and He, assumed 
in cosmological abundance ratio.
The model treats collisional ionization of H$\,{\rm {\scriptstyle I}}$, He$\,{\rm {\scriptstyle I}}$ and
He$\,{\rm {\scriptstyle II}}$, collisional excitation of the same species, and electron 
scattering by Coulomb collisions with thermal electrons. Interactions with He$\,{\rm {\scriptstyle III}}$ 
are neglected and $x_{\rm HeII}=x_{\rm HII}$ is assumed; molecules or 
heavy elements are not accounted for.

A series of analytic expressions for the fraction of the electron energy $E_k$ 
that goes into gas heating ($f_Q$), ionization of species $i$ ($f_{x_i}$), and line 
excitation of hydrogen and helium is provided as function of $E_k$ and gas ionization $x$. 
The fits are valid in the range $E_k=[0.1-3]$~keV and $10^{-4} \leq x \leq 0.95$ with an accuracy
of $\sim$1-2\%. 

The SVS85 model has been implemented in {\tt CRASH4} by importing the values tabulated for $E_k = 100$~eV and 3~keV,  and by interpolating linearly within this energy range to find the energies specified in the source spectra of each {\tt CRASH4} simulation. It should be noted that the values at energies lower than  $E_k = 100$~eV are not provided by the model and they have been evaluated by linear extrapolation down to $E_k =30$~eV, which is a lower limit dictated by the atomic physics of collisions. Also consider that while the interpolation in the range 100~eV-3~keV is justified because of the linear dependence of  $f_{x_i}$ and $f_Q$ on $E_k$ in this energy range, the extrapolation at lower energies is arbitrary and not documented in literature covering the numerical implementation of other codes.

For each electron energy, the binning in $x$ provided by the SVS85 model is maintained, while intermediate values are obtained by linear interpolation. 
We assume that for $x>0.95$  the fractions are the same as for $x=0.95$, while we linearly extrapolate the values of $f_{x_i}$ and $f_Q$ at $x=10^{-4}$ down to $10^{-6}$ (the minimum value of validity for the DG99 model) and assume that $x=0$ for even lower ionization levels.

\begin{figure}
\centering
\includegraphics[angle=-90,width=0.50\textwidth]{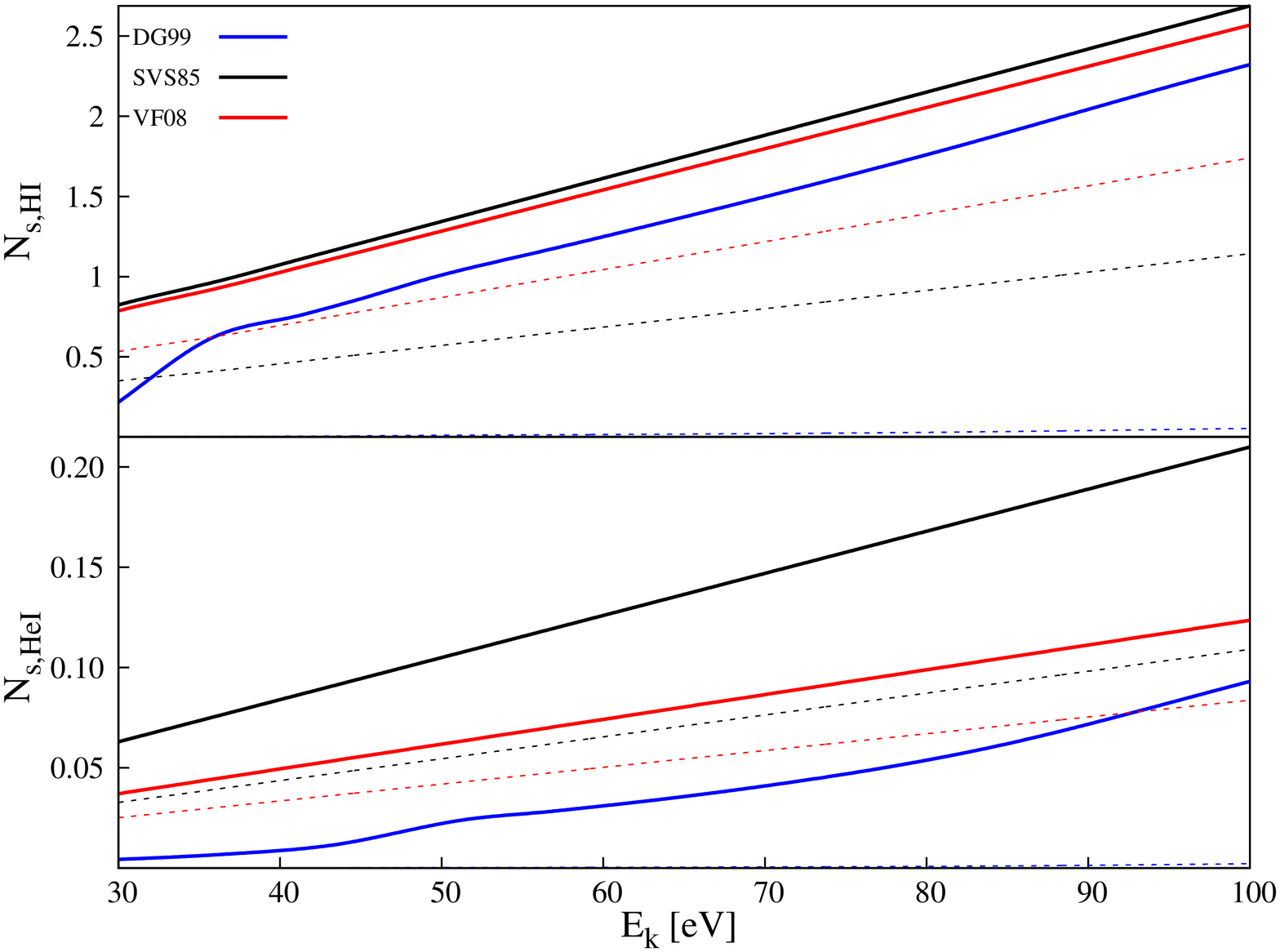}
\vspace{1truecm}
\caption{Average number of collisional ionizations of H$\,{\rm {\scriptstyle I}}$ (top panel) and He$\,{\rm {\scriptstyle I}}$ (bottom panel) as a function of the electron energy $E_k$ in the non-linear regime $E_k < 100$ eV, with $x = 0.1$ (dashed lines) and $x = 10^{-4}$ (solid lines). The DG99, SVS85 and VF08 models are represented by blue, black and red lines, respectively.}
\end{figure}
\begin{figure}
\centering
\includegraphics[angle=-90,width=0.50\textwidth]{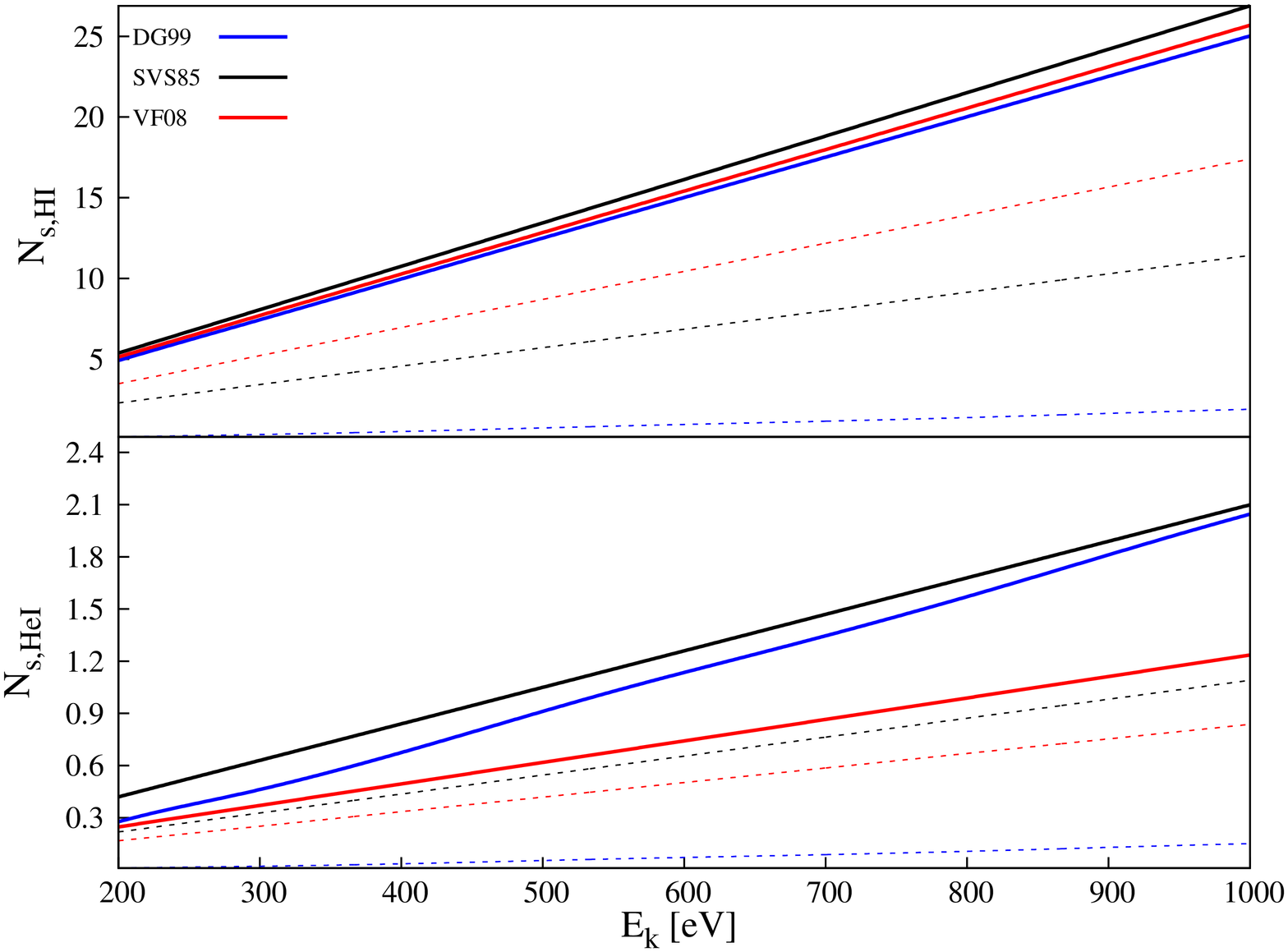}
\vspace{1truecm}
\caption{Average number of collisional ionizations of H$\,{\rm {\scriptstyle I}}$ (top panel) and He$\,{\rm {\scriptstyle I}}$ (bottom panel) as a function of the electron energy $E_k$ in 
         the non-linear regime $E_k > 100$ eV, with 
         $x = 0.1$ (dashed lines) and $x = 10^{-4}$ (solid lines). 
         The DG99, SVS85 and VF08 models are represented by blue, black and red lines, respectively.   
        }
\end{figure}

\subsection{VF08 model} 

The VF08 model provides one of the most advanced and updated 
implementations of atomic data in the literature and fitting trends valid for $E_k>3$~keV. 
The adoption of VF08 will allow then a future integration 
of secondary effects from the hard X-ray and gamma spectral bands \citep{2010MNRAS.404.1569V}.
Following the method used by SVS85, this model includes the most recent atomic 
data, accurate cross sections for collisional ionization and excitations from 
electron impacts, electron-electron collisions, free-free interactions and gas 
recombinations. Together with 
the inclusion of the generally neglected two-photon forbidden transition $2s \rightarrow 1s$, 
processes producing continuum photons (i.e. gas recombinations, Bremsstrahlung 
free-free and ion-electron interactions) are also accounted for. 
As SVS85, they assume a gas made of H and He in cosmological proportions and $x_{\rm HeII}=x_{\rm HII}$, neglecting doubly ionized helium.
Values for the fraction of energy going into photons with energy below 10.2~eV are provided as well. 

Fitting functions for $f_{x_i}$ and $f_Q$ are provided in the range $E_k$=[3-10]~keV, with an accuracy better than 3.5\%. The validity range in ionization fraction is $10^{-4}\leq x \leq 0.99$.
For $x<0.1$ the functions are very similar to those provided by SVS85, while differences as large as 30\% are found for higher ionization fractions. 

The inclusion of VF08 in \texttt{CRASH4} has been done as the one of SVS85, i.e.  by importing the values tabulated for $E_k = 10$ keV, and evaluating the values at lower energies by linear extrapolation down to $E_k =30$~eV. The extension of the tables to a wider range of ionization fraction has been done as for SVS85.

\paragraph*{}
In the paragraphs below we summarise the behaviour of the three models by showing the average 
number of collisional ionizations, $N_{s,i}$, as function of the photo-electron energy $E_k$ in 
the non-linear (Figure A1) and linear regime (Figure A2). In both figures we only show
the ionizations of H$\,{\rm {\scriptstyle I}}$  and 
He$\,{\rm {\scriptstyle I}}$ because only the DG99 model provides values for   He$\,{\rm {\scriptstyle II}}$.
We show curves for $x=10^{-4}$ and 0.1, which are the minimum and maximum ionization fraction shared by the three models.
For $E_k<100$~eV DG99 is the only accurate model, while the curves from SVS85 and VF08 are obtained by linear extrapolation of results at $E_k=100$~eV and 3~keV, respectively. From Figure~A1 it is clear that the number of collisional ionizations is largely overestimated when a linear extrapolation is applied, and that the discrepancy with the DG99 model increases with the ionization fraction. In fact, while for $x\gsim 0.1$ hardly any ionization occurs in the DG99 model, this is not the case for SVS85 and VF08. The above considerations apply both for H$\,{\rm {\scriptstyle I}}$ and He$\,{\rm {\scriptstyle I}}$.
Substantial differences exist also in the linear regime shown in Figure~A2. While in a quasi-neutral medium (solid lines) the agreement between the models is extremely good (at least for H$\,{\rm {\scriptstyle I}}$), it rapidly worsens with  increasing ionization fractions, with SVS85 and VF08 finding larger $N_{\rm s,HI}$ and $N_{\rm s,HeI}$, as in the non linear regime.

\section{Impact of secondary ionization models on H$\,{\rm {\scriptstyle II}}$ regions}

To understand how the choice of a specific secondary ionization model impacts 
the final properties of the H$\,{\rm {\scriptstyle II}}$ regions (see Figure 2), we ran 
the same test of Section 3.2 by adopting the SVS85 and VF08 models. In addition to the 
standard ionization rate $\dot{N}_{\gamma,0} = 1.36 \times 10^{56}$ photons s$^{-1}$ we also considered 
the case with $\dot{N}_{\gamma,1} = 1.36 \times 10^{57}$ photons s$^{-1}$.  

The results of this comparison are summarized in Figure B1 for a case in which 
the contribution from X-rays and secondary ionization is included.
First note that for the simple configuration of this test the results provided by 
SVS85 and VF08 are indistinguishable in the case with $\dot{N}_{\gamma,0}$, while few small differences 
in the hydrogen and helium profiles are found in the remaining cases.

\begin{figure}
\centering
\includegraphics[angle=-90,width=0.50\textwidth]{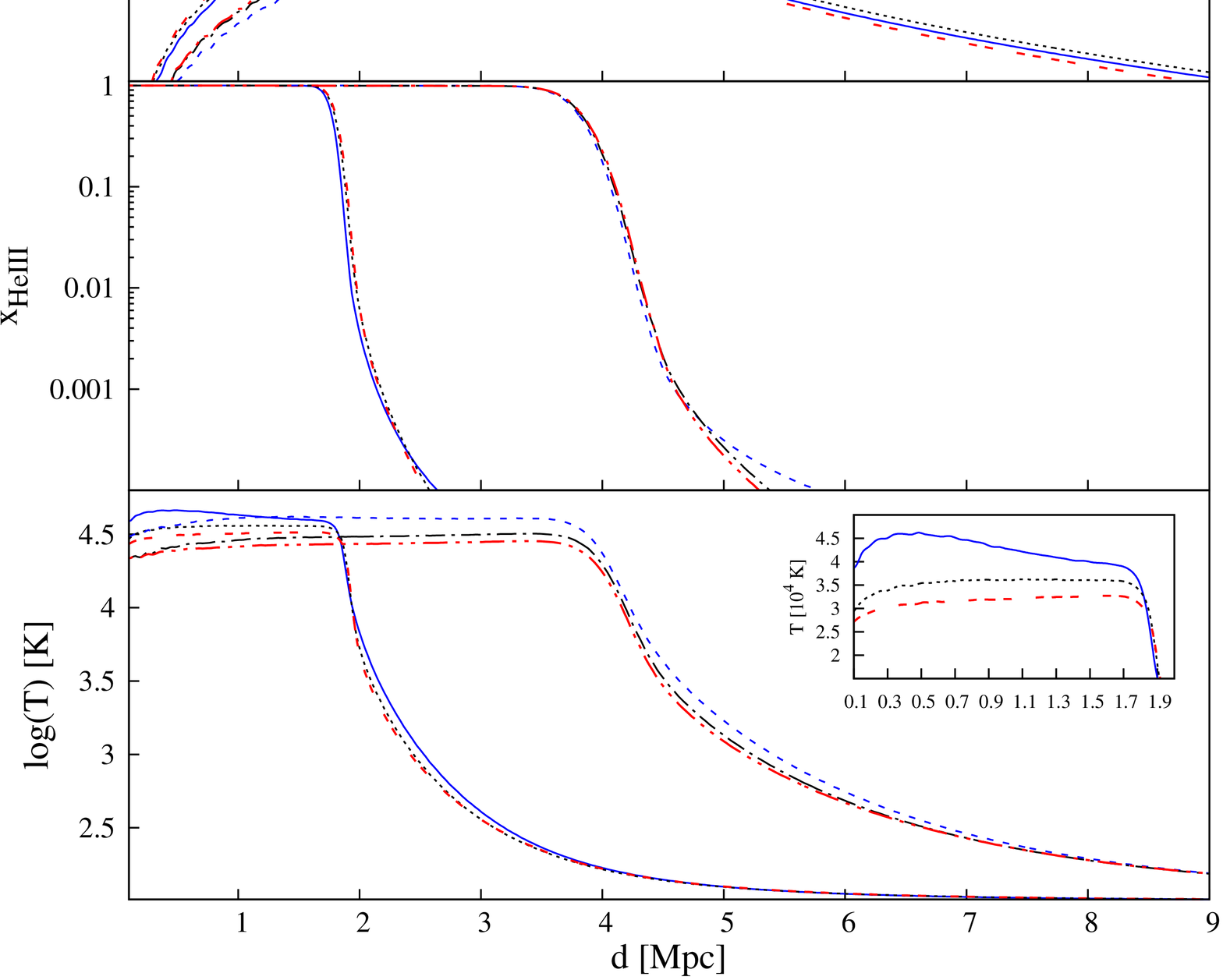}
\vspace{1truecm}
\caption{Radial profiles of the  Str\"omgren sphere created in a cosmological mixture of 
         H and He at the simulation time $t_i= 6\times10^7$ yrs. 
         The distance $d$ from the source is shown in physical Mpc. 
         In all the panels
         the profiles are obtained using the model: DG99 (blue lines), 
         SVS85 (black) and VF08 (red).
         For each simulation set two cases are considered: $\dot{N}_{\gamma,0} = 1.36 \times 10^{56}$
         photons s$^{-1}$ and $\dot{N}_{\gamma,1} = 1.36 \times 10^{57}$ photons s$^{-1}$. 
         From the top to the bottom the panels refer to the profiles of 
         $x_{\textrm{HII}}$, $x_{\textrm{HeII}}$, $x_{\textrm{HeIII}}$ and log$(T)$ [K].
         The small box in the bottom panel shows a zoom of the gas temperature (in linear
         scale of $10^4$ K) in the fully ionized region 0.1 Mpc $< d <2.0$ Mpc, physical.
         }         
\end{figure}

Second, while a discrepancy is found between DG99 and the other models and it increases at higher ionization rates, DG99 always predicts a radius of the I-front smaller by 13\%. These differences are due to the different behavior of the models in the limit $x_e \rightarrow 0$ (see Appendix A). Other differences in the $x_{\rm HeII}$ and $x_{\rm HeIII}$ profiles are also present although less evident.
 
The highest discrepancy though is observed in the temperature (bottom panel and small box inside it), which is systematically higher in the DG99 model. 
These differences are due not only to the linear extrapolation of models SVS85 and VF08 at energies below their valodity limit, but in the way the heating mechanism is modelled. SVS85 and VF08 tend in fact to 
over-estimate the ionization reducing the contribution to the photo-heating (see Figures A1 and A2 in Appendix A) but also to distribute more energy to excitations instead of gas photo-heating.

\section{Effects of initial conditions}

Here we show a series of ideal tests repeated in a range of gas number densities, 
and non neutral initial conditions.  The reference secondary ionization model adopted for runs with varying number density is DG99, while VF08 is adopted for tests with varying initial ionization fractions to allow a direct comparison with the results in Appendix B and Section 4.

\begin{figure}
\centering
\includegraphics[angle=-90,width=0.50\textwidth]{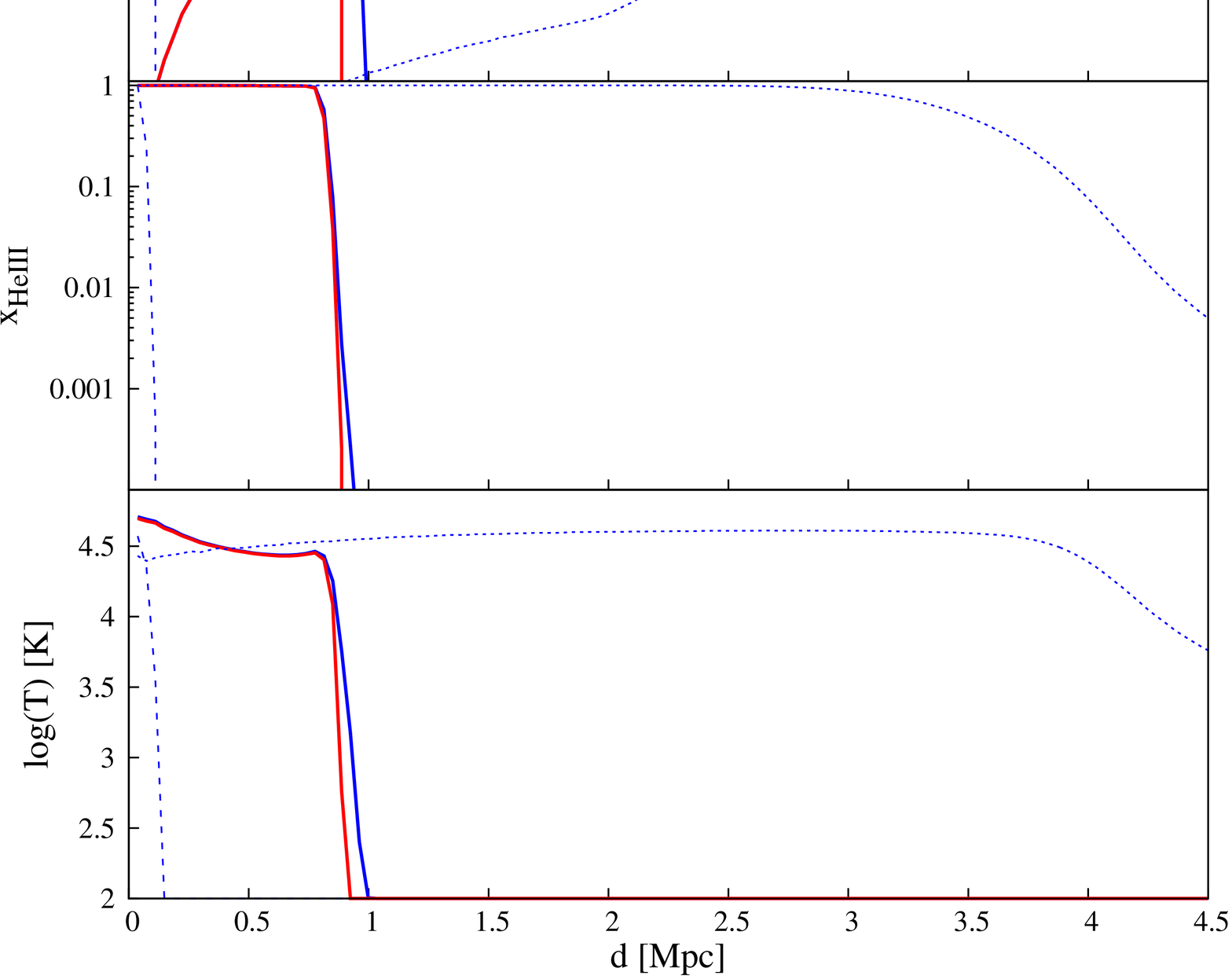}
\vspace{1truecm}
\caption{Radial profiles of the Str\"omgren sphere created by a QSO-like source  in
         cosmological mixture of 
         H and He by adopting the DG99 model and different values of the gas number 
         density: $n_{gas}=10^{-1}$~cm$^{-3}$ (dashed lines), 
         $10^{-3}$~cm$^{-3}$ (solid), 
         and $10^{-5}$~cm$^{-3}$ (dotted). 
         The lines refer to the simulation time 
         in all the configurations $t_i=6\times10^7$ yrs. 
         The distance $d$ from the source is shown in physical Mpc. In all the panels
         the profiles referring to the (UV+X) case  are represented as blue 
         lines. Only for the case $n_{gas}=10^{-3}$~cm$^{-3}$ we show the profile with UV radiation only(red lines). 
         From the top to the bottom the panels refer to the profiles of 
         $x_{\textrm{HII}}$, $x_{\textrm{HeII}}$, $x_{\textrm{HeIII}}$ and log$(T)$ [K].
         }         
\end{figure}

Here we show ideal tests adopting the same parameters as in Section 4 but for the gas density which is set up as uniform in space.
As always the convergence of the solutions has been verified with an identical simulation increasing the number of packets by one order of magnitude. 
We varied the gas number density in the range $n_{gas} \in [10^{-1}, 10^{-5}]$ cm$^{-3}$;  some representative profiles are shown in Figure C1 with the same color convention used in Section 3. Here different line-styles correspond to different 
values of the gas number density: $n_{gas}=10^{-1}$~cm$^{-3}$ in dashed, $n_{gas}=10^{-3}$~cm$^{-3}$ in solid, and finally $n_{gas}=10^{-5}$~cm$^{-3}$ in dotted. As always we show the profiles in physical distance $d$.   

First note how the profiles of the Str\"omgren spheres are really sensitive to the gas number density both in the I-front positions and the extended shells at low ionization found in the reference case (solid lines).
For example in the top panel we show how the hydrogen I-fronts of the spheres change from $d \sim 0.15$ Mpc to $d \sim 0.95$ Mpc, and finally $d > 3.7$ Mpc, when the gas number density changes by one order of 
magnitude. The external shell increases its extension accordingly (compare the blue lines) and escapes 
the domain boundaries when $n_{gas}=10^{-5}$~cm$^{-3}$. By comparing the blue and red lines we note also how the X-rays 
band becomes progressively more effective in creating a diffuse profile when the gas number density decreases. 
Similar considerations can be applied at the ions of helium in the middle and bottom panels. In particular note how easy it is to maintain the helium fully ionized at large distances when the number density reached $n_{gas}=10^{-5}$~cm$^{-3}$. 

The temperature profiles are impacted as well, as 
shown in the bottom panel. The entire cube is set up at photo-ionization temperature in the lowest density case while in the case $n_{gas}=10^{-3}$~cm$^{-3}$ the gas temperature is maintained at $T>10^{3}$ K at distances larger than $d \sim 0.9$ physical Mpc only by the X-rays contribution. The column density created by $n_{gas}=10^{-1}$~cm$^{-3}$ on the other hand is sufficient to completely stop the soft X-ray flux by 0.15 Mpc from the QSO source.

In the paragraphs below we study the dependence of our results when an initial ionization $x_0$ of the gas is assumed. For this test we adopt the gas number density of the reference case $n_{gas}=9.59 \times 10^{-5}$ cm$^{-3}$ and assume an initial ionization $x_{0,\textrm{HII}}=x_{0,\textrm{HeII}}=x_0=$0.01, 0.1 and 0.5, while $x_{0,\textrm{HeIII}}=0$.

When a pre-ionization  $x_0 < 0.1$ is set up, all the species rapidly increase their ionization in the external, low ionization region ($d > 1.5 $ Mpc), as evident by comparing all the lines with the dashed-dotted one. This is because an initial ionization decreases the gas optical depth in both H$\,{\rm {\scriptstyle II}}$ and He$\,{\rm {\scriptstyle II}}$ and then the profiles of He$\,{\rm {\scriptstyle III}}$ change as well. 
Longer tails in the temperature profiles are also created (see bottom panel) by inducing gas temperatures higher than $T=1000$~K at $d > 2.0$ Mpc.

Smaller effects are found instead on the ionization I-fronts of the Str\"omgren spheres which show a noticeable shift only for $x_0 > 0.1$. Also note that a pre-ionization condition has a substantial impact on the temperature of the fully ionized gas which reaches a lower photo-ionization equilibrium temperature (compare dotted curve with dashed-dotted in the bottom panel). 

As final remark, we point out that tests performed by mixing initial ionization $x_0 \sim 0.01$  and temperatures $T_0 \sim 10^{3}$ K have been performed as well during the test of \texttt{CRASH4} without finding remarkable differences with the results shown in this section. 

\section{Infinite speed of light effects on time evolution}

Here we repeat Tests 3.1 and 3.2 with longer simulation duration to capture the late evolution of the HII region. To find the equilibrium configuration in a gas of $n_{\textrm{gas}}=9.59 \times 10^{-5}$ cm$^{-3}$ we should set up a final time  $t_f = 4 \times t_{rec} \sim 4\times 10^9$~yrs, as prescribed in a similar test of the Cosmological Radiative Transfer Comparison Project \citep{2006MNRAS.371.1057I}. Note that while $t_f \sim 10^9$~yrs is still consistent with extreme low-redshift QSO lifetimes \citep{2001ApJ...547...12M}, our simulation would reach $z\sim 1.34$ when the assumption that the bubble tail expands in a neutral gas (i.e. isolated QSO) is unrealistic. We have thus chosen to end the simulation at $t_f = 10^8$~yrs, which corresponds to $z\sim 6.45$. This time is sufficiently long to follow the late evolution of the tails in our idealized set-up. The box size is $110h^{-1}$ Mpc comoving but with  the same grid resolution as in Section 3.1.

\begin{figure}
\centering
\includegraphics[angle=-90,width=0.50\textwidth]{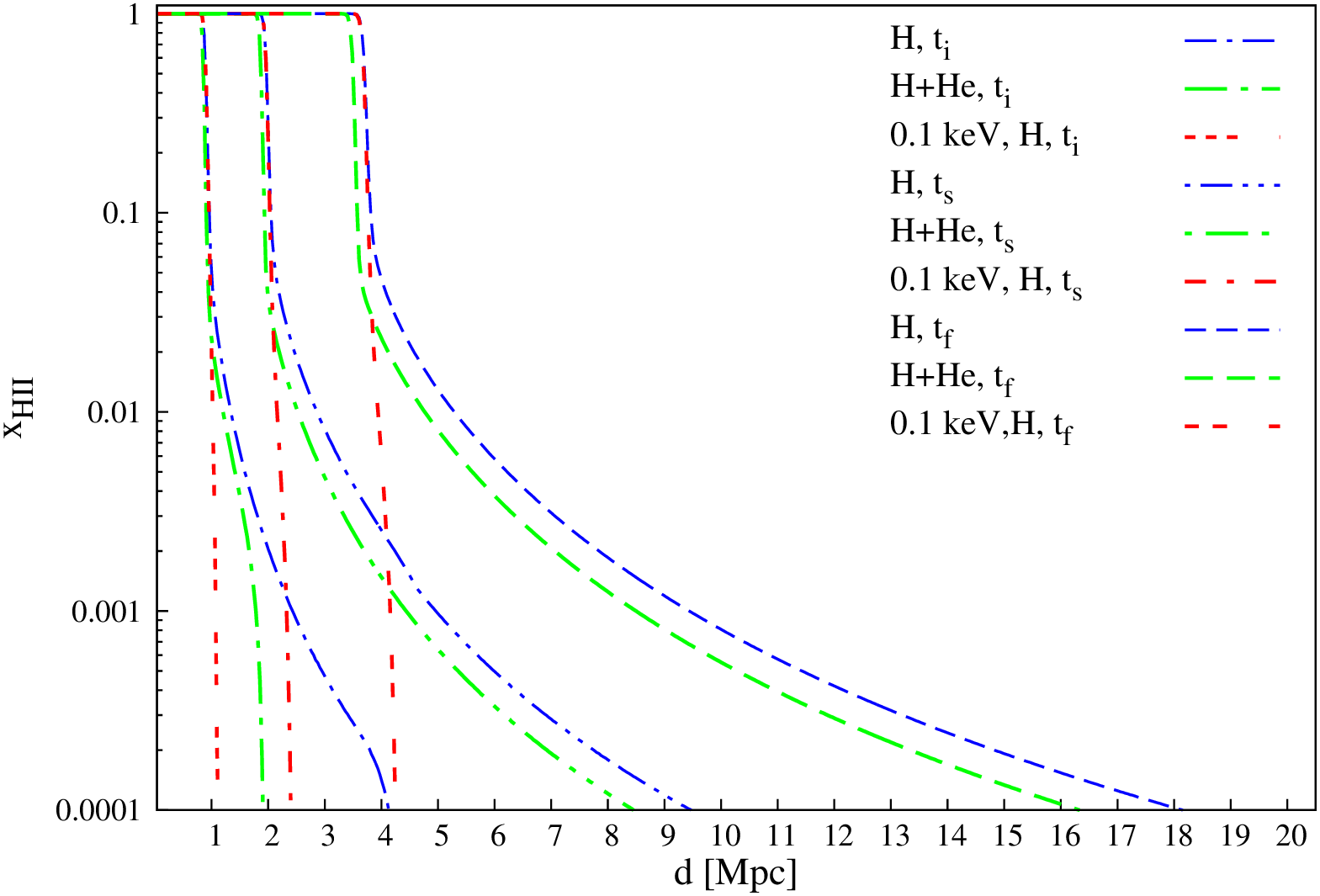}
\vspace{1truecm}
\caption{Radial profiles of $x_{\textrm{HII}}$ of the Str\"omgren sphere created in a gas made of 
         neutral atomic hydrogen only (blue/red lines) and H+He (green) at different simulation 
         times: $t_i=2\times10^6$~yrs, $t_s=2\times10^7$~yrs, $t_f=10^8$~yrs. 
         The secondary ionization model adopted in the simulation is DG99. 
         The distance $d$ from the source is shown in physical Mpc. 
         Blue and green lines refer to a spectrum cut at $E_{\textrm{max}}= 3$~keV, while  
         red lines refer to $E_{\textrm{max}}= 0.1$~keV.
         }
\end{figure}

\begin{table*}
\begin{centering}
\begin{tabular}{|c|c|c|c|c|c|c| c}
\hline
$t$~[yrs] & $z$ & $R_c$~[Mpc] & $R_s$~[Mpc] & $v_{f,c}$ [$c$] & $d_{\textrm{max}, 0.001}$~[Mpc](H/H+He) &  $d_{\textrm{max}, 0.0001}$~[Mpc](H/H+He) & $d_{c,\textrm{vac}}$~[Mpc]\tabularnewline
\hline
\hline
$2\times10^6$ & $7.07$ & $0.850$  & $0.891$ & $1.24$ & 2.40/1.77 & 4.10/1.88 & 0.6 \tabularnewline
\hline
$4\times10^6$ & $7.06$ & $1.072$  & $1.123$ & $0.78$ & 3.0/2.0 & 4.44/2.52 &  1.23 \tabularnewline
\hline
$6\times10^6$ & $7.04$ & $1.259$  & $1.285$ & $0.61$ & 3.45/2.45 & 5.34/4.52 & 1.85 \tabularnewline
\hline
$8\times10^6$ & $7.03$ & $1.373$  & $1.415$ & $0.51$ & 3.75/2.93 & 6.32/5.50 & 2.45 \tabularnewline
\hline
$10^7$ & $7.01$ & $1.487$  & $1.524$ & $0.43$ & 4.02/3.31 & 7.04/6.22 & 3.07 \tabularnewline
\hline
$2\times10^7$ & $6.94$ & $1.912$  & $1.920$ & $0.28$ & 4.95/4.43 & 9.46/8.45 & 6.14 \tabularnewline
\hline
$4\times10^7$ & $6.82$ & $2.478$  & $2.419$ & $0.18$ & 6.45/5.80 & 12.40/11.14 & 12.27 \tabularnewline
\hline
$6\times10^7$ & $6.69$ & $2.906$  & $2.769$ & $0.14$ & 7.56/6.83 & 14.58/13.11 & 18.50 \tabularnewline
\hline
$10^8$ & $6.45$ & $3.64$  & $3.283$ & $0.11$ & 9.41/8.49 & 18.14/16.30 & 30.70 \tabularnewline
\hline
\end{tabular}
\par\end{centering}
\caption{\label{tab:fCGalaxies} Time evolution of the $x_{\textrm{HII}}$ profiles. $t$ is the simulation time in years, $z$ is the redshift, $R_c$ is the I-front position computed by {\tt CRASH4} and $R_s$ is the I-front of the semi-analytic estimate in formula (3), both in proper Mpc. $v_{f,c}$ is the I-front speed computed from $R_c$, in units of $c\sim 3\times 10^{10}$~cm s$^{-1}$. $d_{\textrm{max}, 0.001}$  and $d_{\textrm{max}, 0.0001}$ are the maximum extension of the $x_{\textrm{HII}}$ profiles at $x_{\textrm{HII}} > 0.001$ and $x_{\textrm{HII}} > 0.0001$, respectively. The H/H+He nomenclature refers to the H only and H+He cases. As a reference, the last column shows the maximum distance traveled by light in vacuum at fixed time ($d_{c,\textrm{vac}}$).}
\end{table*}

Figure D1 shows the spherically averaged radial profiles of $x_{\textrm{HII}}$ at different times for both 0.1~keV and 3~keV cases; by comparing them the contribution of X-rays to the tails becomes evident\footnote{We remind here that when $E_{\gamma}> 100$~eV the efficiency of secondary processes is quasi-linear. See Appendix A for more details.}. The more realistic case in which helium is included is also shown by green lines. Table D1 complements Figure D1 by providing more details on the fronts obtained with the cosmic web discussed in Section 4 at early and late times: $R_c$ is the value of the I-front position computed by {\tt CRASH4}, $R_s$ the value inferred by equation (3) and $v_{f,c}$ is the front speed computed for $R_c$, in units of physical speed of light $c\sim 3\times10^{10}$~cm s$^{-1}$. The maximum tail extension at assigned time and ionization fraction ($d_{max,x_{\textrm{HII}}}$) is computed for  $x_{\textrm{HII}} = 0.001, 0.0001$ and can be compared with the maximum distance traveled by light in vacuum ($d_{c, \textrm{vac}}$). Note that $x_{\textrm{HII}} = 0.0001$ is the lowest Monte Carlo convergent value found in our simulation by sampling the source with $2\times10^8$ photon packets.

The extension of the tails is overestimated at short times as they are present beyond the maximum distance traveled by radiation in vacuum. This error is due to the infinite speed of light approximation which creates an excess of high energy photons beyond the I-front and induces an early growth of the external regions. As shown in the table, this effect becomes progressively less evident in time and disappears at $t=6\times10^7$~yrs. The effective impact on the time evolution of the tail cannot be assessed with a single rule of thumb, as it depends both on the source properties (e.g. it is higher in the early evolution of extremely bright sources), gas number density and the efficiency of secondary electrons. In conclusion, its impact has to be verified depending on the problem at hand.

\end{appendix}

\label{lastpage}
\end{document}